\documentclass[twocolumn,twocolappendix]{aastex63}

\newcommand{\mjup}{$M_{\rm Jup}$}
\newcommand{\teff}{$T_{\rm eff}$}
\newcommand{\kms}{km s$^{-1}$}

\usepackage{color}
\usepackage{float}
\shorttitle{Brown Dwarf Co-Moving Companions}
\shortauthors{Rothermich et al.}
\graphicspath{{./}{figures/}}

\begin{document}

\title{89 New Ultracool Dwarf Co-Moving Companions Identified With The Backyard Worlds: Planet 9 Citizen Science Project}

\author[0000-0003-4083-9962]{Austin Rothermich}
\affiliation{Department of Astrophysics, American Museum of Natural History, Central Park West at 79th Street, NY 10024, USA}
\affiliation{Department of Physics, Graduate Center, City University of New York, 365 5th Ave., New York, NY 10016, USA}
\affiliation{Department of Physics and Astronomy, Hunter College, City University of New York, 695 Park Avenue, New York, NY, 10065, USA}
\affiliation{Backyard Worlds: Planet 9}

\author[0000-0001-6251-0573]{Jacqueline K. Faherty}
\affiliation{Department of Astrophysics, American Museum of Natural History, Central Park West at 79th Street, NY 10024, USA}
\affiliation{Backyard Worlds: Planet 9}

\author[0000-0001-8170-7072]{Daniella Bardalez-Gagliuffi}
\affiliation{Department of Astrophysics, American Museum of Natural History, Central Park West at 79th Street, NY 10024, USA}
\affiliation{Backyard Worlds: Planet 9}

\author[0000-0002-6294-5937]{Adam C. Schneider}
\affiliation{US Naval Observatory, Flagstaff Station, 10391 West Naval Observatory Road, Flagstaff, AZ 86005, USA}
\affiliation{Backyard Worlds: Planet 9}

\author[0000-0003-4269-260X]{J. Davy Kirkpatrick}
\affiliation{IPAC, Mail Code 100-22, Caltech, 1200 E. California Blvd., Pasadena, CA 91125, USA}
\affiliation{Backyard Worlds: Planet 9}

\author[0000-0002-1125-7384]{Aaron M. Meisner}
\affiliation{NSF's National Optical-Infrared Astronomy Research Laboratory, 950 N. Cherry Ave., Tucson, AZ 85719, USA}
\affiliation{Backyard Worlds: Planet 9}

\author[0000-0002-6523-9536]{Adam J.\ Burgasser}
\affil{Department of Astronomy \& Astrophysics, University of California, San Diego, 9500 Gilman Drive, La Jolla, CA 92093, USA}
\affiliation{Backyard Worlds: Planet 9}

\author[0000-0002-2387-5489]{Marc Kuchner}
\affiliation{Exoplanets and Stellar Astrophysics Laboratory, NASA Goddard Space Flight Center, 8800 Greenbelt Road, Greenbelt, MD 20771, USA}
\affiliation{Backyard Worlds: Planet 9}

\author{Katelyn Allers}
\affil{Physics and Astronomy Department, Bucknell University, 701 Moore Ave, Lewisburg, PA 17837, USA}

\author[0000-0002-2592-9612	]{Jonathan Gagn\'{e}}
\affil{ Plan\`{e}tarium Rio Tinto Alcan, 4801 Pierre-de Coubertin Ave, Montreal, Quebec H1V 3V4, Canada}
\affiliation{Backyard Worlds: Planet 9}

\author[0000-0001-7896-5791]{Dan Caselden}
\affiliation{Department of Astrophysics, American Museum of Natural History, Central Park West at 79th Street, NY 10024, USA}
\affiliation{Backyard Worlds: Planet 9}

\author[0000-0002-2682-0790]{Emily Calamari}
\affiliation{Department of Astrophysics, American Museum of Natural History, Central Park West at 79th Street, NY 10024, USA}
\affiliation{Department of Physics, Graduate Center, City University of New York, 365 5th Ave., New York, NY 10016, USA}

\author[0000-0001-9482-7794]{Mark Popinchalk}
\affiliation{Department of Astrophysics, American Museum of Natural History, Central Park West at 79th Street, NY 10024, USA}
\affiliation{Department of Physics, Graduate Center, City University of New York, 365 5th Ave., New York, NY 10016, USA}
\affiliation{Department of Physics and Astronomy, Hunter College, City University of New York, 695 Park Avenue, New York, NY, 10065, USA}

\author[0000-0002-2011-4924]{Genaro Su\'{a}rez}
\affiliation{Department of Astrophysics, American Museum of Natural History, Central Park West at 79th Street, NY 10024, USA}

\author[0000-0003-0398-639X]{Roman Gerasimov}
\affil{Center for Astrophysics and Space Science, University of California, San Diego, 9500 Gilman Drive, La Jolla, CA 92093, USA}
\affil{Department of Physics \& Astronomy, University of Notre Dame, Notre Dame, IN 46556, USA}

\author[0000-0003-2094-9128]{Christian Aganze}
\affil{Center for Astrophysics and Space Sciences, University of California, San Diego, 9500 Gilman Drive, Mail Code 0424, La Jolla, CA 92093, USA}
\affil{Department of Physics, Stanford University, Stanford CA 94305, USA}

\author[0000-0002-1420-1837]{Emma Softich}
\affil{Department of Astronomy \& Astrophysics, University of California, San Diego, 9500 Gilman Drive, La Jolla, CA 92093, USA}

\author[0000-0002-5370-7494]{Chin-Chun Hsu}
\affil{Center for Interdisciplinary Exploration and Research in Astrophysics (CIERA), Northwestern University,
1800 Sherman, Evanston, IL, 60201, USA}

\author{Preethi Karpoor}
\affil{Department of Astronomy \& Astrophysics, University of California, San Diego, 9500 Gilman Drive, La Jolla, CA 92093, USA}

\author[0000-0002-9807-5435]{Christopher A.\ Theissen}
\affil{Department of Astronomy \& Astrophysics, University of California, San Diego, 9500 Gilman Drive, La Jolla, CA 92093, USA}

\author[0000-0002-5376-3883]{Jon Rees}
\affil{UCO/Lick Observatory, 7281 Mount Hamilton Road, Mount Hamilton, CA 95140, USA}

\author{Rosario Cecilio-Flores-Elie}
\affiliation{Department of Physics, Graduate Center, City University of New York, 365 5th Ave., New York, NY 10016, USA}

\author{Michael C.\ Cushing}
\affiliation{Ritter Astrophysical Research Center, Department of Physics and Astronomy, University of Toledo, 2801 W. Bancroft Street, Toledo, OH 43606, USA}

\author[0000-0001-7519-1700]{Federico Marocco}
\affil{IPAC, Mail Code 100-22, Caltech, 1200 E. California Blvd., Pasadena, CA 91125, USA}
\affiliation{Backyard Worlds: Planet 9}

\author[0000-0003-2478-0120]{Sarah Casewell}
\affil{School of Physics and Astronomy, University of Leicester, University Road, Leicester, LE1 7RH, UK}
\affiliation{Backyard Worlds: Planet 9}

\author[0000-0003-2235-761X]{Thomas P. Bickle}
\affiliation{School of Physical Sciences, The Open University, Milton Keynes, MK7 6AA, UK}
\affiliation{Backyard Worlds: Planet 9}

\author{Les Hamlet}
\affiliation{Backyard Worlds: Planet 9}

\author{Michaela B.\ Allen}
\affiliation{NASA Goddard Space Flight Center, Exoplanets and Stellar Astrophysics Laboratory, Code 667, Greenbelt, MD 20771, USA}

\author{Paul Beaulieu}
\affiliation{Backyard Worlds: Planet 9}

\author[0000-0002-7630-1243]{Guillaume Colin}
\affiliation{Backyard Worlds: Planet 9}

\author[0000-0002-1044-1112]{Jean Marc Gantier}
\affiliation{Backyard Worlds: Planet 9}

\author[0000-0002-8960-4964]{Leopold Gramaize}
\affiliation{Backyard Worlds: Planet 9}

\author[0000-0002-4175-295X]{Peter Jalowiczor}
\affiliation{Backyard Worlds: Planet 9}

\author[0000-0003-4905-1370]{Martin Kabatnik}
\affiliation{Backyard Worlds: Planet 9}

\author[0000-0001-8662-1622]{Frank Kiwy}
\affiliation{Backyard Worlds: Planet 9}

\author{David W. Martin}
\affiliation{Backyard Worlds: Planet 9}

\author{Billy Pendrill}
\affiliation{Backyard Worlds: Planet 9}

\author[0000-0001-9692-7908]{Ben Pumphrey}
\affiliation{Backyard Worlds: Planet 9}

\author[0000-0003-4864-5484]{Arttu Sainio}
\affiliation{Backyard Worlds: Planet 9}

\author[0000-0002-7587-7195]{Jorg Schumann}
\affiliation{Backyard Worlds: Planet 9}

\author[0000-0003-4714-3829]{Nikolaj Stevnbak}
\affiliation{Backyard Worlds: Planet 9}

\author[0000-0003-3162-3350]{Guoyou Sun}
\affiliation{Backyard Worlds: Planet 9}

\author{Christopher Tanner}
\affiliation{Backyard Worlds: Planet 9}

\author{Vinod Thakur}
\affiliation{Backyard Worlds: Planet 9}

\author[0000-0001-5284-9231]{Melina Thevenot}
\affiliation{Backyard Worlds: Planet 9}

\author{Zbigniew Wedracki}
\affiliation{Backyard Worlds: Planet 9}



\begin{abstract}

We report the identification of 89 new systems containing ultracool dwarf companions to main sequence stars and white dwarfs, using the citizen science project Backyard Worlds: Planet 9 and cross-reference between Gaia and CatWISE2020. 
Thirty-two of these companions and thirty-three host stars were followed up with spectroscopic observations, with companion spectral types ranging from M7-T9 and host spectral types ranging from G2-M9. These systems exhibit diverse characteristics, from young to old ages, blue to very red spectral morphologies, potential membership to known young moving groups, and evidence of spectral binarity in 9 companions. Twenty of the host stars in our sample show evidence for higher order multiplicity, with an additional 11 host stars being resolved binaries themselves. We compare this sample's characteristics with those of the known stellar binary and exoplanet populations, and find our sample begins to fill in the gap between directly imaged exoplanets and stellary binaries on mass ratio-binding energy plots. With this study, we increase the population of ultracool dwarf companions to FGK stars by $\sim$42\%, and more than triple the known population of ultracool dwarf companions with separations larger than 1,000 au, providing excellent targets for future atmospheric retrievals. 

\end{abstract}



\section{Introduction} \label{sec:intro}
Ultracool dwarfs, which consist of objects of effective temperatures (\teff) below $\sim$2,700 K, include late-M spectral types and sub-stellar mass objects known as brown dwarfs. Brown dwarfs bridge the mass regimes of stars and giant exoplanets, with masses below the hydrogen burning limit ($\sim$75 \mjup) and above the deuterium burning limit ($\sim$13 \mjup) \citep{1963ApJ...137.1121K,1963PThPh..30..460H, 1997ApJ...491..856B}. Brown dwarfs likely form through direct collapse and fragmentation of their natal molecular cloud \citep{2002MNRAS.332L..65B, 2003MNRAS.339..577B, 2005MNRAS.356.1201B, 2009MNRAS.392..590B, 2012MNRAS.419.3115B} and therefore represent a low mass extension to star formation. The growing number of isolated planetary mass sources discovered (for example, \citealt{2015ApJS..219...33G}) suggests this process is able to produce objects that overlap with the mass regime of giant exoplanets. Many brown dwarf atmospheres also have temperatures comparable to those found in exoplanets ($\sim$250--3000 K; 
\citealt{2005ARA&A..43..195K}), 
meaning both populations share atmospheric chemistry. While the direct study of exoplanets is difficult, due in part to the contamination from their host star's light, many brown dwarfs are found without a contaminating bright host, providing easier targets for spectroscopic observations. Brown dwarfs are therefore ideal exoplanet analogs for the testing and constraining of atmospheric models \citep{2016ApJS..225...10F}. 

Since the first confirmed discoveries \citep{1995Natur.377..129R,1995Natur.378..463N,1995Sci...270.1478O}, numerous brown dwarfs have been identified (see \citealt{2005ARA&A..43..195K}). Characterizing brown dwarfs remains difficult though, due to a degeneracy between mass, temperature, and age \citep{2001RvMP...73..719B}. The large number of different molecular species which form in brown dwarf atmospheres presents another challenge, as it can be difficult to disentangle the physical properties that dictate atmospheric chemistry and thereby sculpt their highly structured spectra \citep{2005ARA&A..43..195K, 2014A&ARv..22...80H}. These factors make it a challenge to fully characterize an isolated brown dwarf.
 
Brown dwarfs as co-moving companions to more easily characterized stars adopt many of the properties of their hosts. Assuming the system to be coeval and having formed from the same molecular cloud, properties like age and metallicity that are measurable  in the primary are expected to apply to the companion as well. When in wide orbits (e.g. $>$100 AU), brown dwarf companions within a few tens of parsecs can be fully resolved, allowing them to be observed without interference from their host star, making these valuable ``benchmark'' systems \citep[e.g.,][]{2006MNRAS.368.1281P, 2014ApJ...792..119D, 2017MNRAS.470.4885M}. 
Wide stellar companions have been used to calibrate the ages of main sequence stars \citep{2011A&A...531A...7G,2019ApJ...870....9F}, as well as spectroscopic metallicity relations of M dwarfs \citep{2010ApJ...720L.113R,2013ApJ...779..188M,2014AJ....147..160M}. The use of benchmarks even extends to groups of related objects, like young moving groups, which are co-evolving and share common chemical and kinematic histories \citep[e.g.,][]{2004ARA&A..42..685Z}.    

In addition to serving as benchmarks for the calibration of atmospheric models, brown dwarf companions present the opportunity to investigate the formation pathways of substellar objects. Stellar abundance ratios, such as C/O, have been proposed as a tracer of formation mechanisms \citep[e.g.,][]{2016ApJ...832...41M, 2017ApJ...838L...9E}. If the C/O ratios between the host and its companion are within statistical agreement it is likely they formed simultaneously through direct collapse, whereas disagreement in C/O ratio may be indicative of a core accretion scenario for the substellar companion \citep{2011ApJ...743L..16O, 2012ApJ...758...36M}. However, there is evidence that this assumption may not hold true, as some brown dwarfs which likely formed via direct collapse appear to have elevated C/O ratios \citep{2022ApJ...940..164C}. Recent works have begun using atmospheric retrievals of brown dwarfs to derive abundances, temperature-pressure profiles, and other fundamental parameters \citep[e.g.,][]{2020ApJ...905...46G, 2021MNRAS.506.1944B, 2022ApJ...936...44Z, 2022ApJ...930..136L, 2023ApJ...944..138V}. For benchmark systems, the ability to compare the derived abundances with those of the host star can ground the analysis of brown dwarf atmospheric chemistry with external empirical information \citep{2022ApJ...940..164C,2022AJ....163..189W,2023AJ....166...85H,2023MNRAS.521.5761G}. 

In this paper, we present the discovery of 89 low mass star and brown dwarf (together referred to as ultracool dwarfs) benchmark systems, with spectral types between M7 and T9, identified through the citizen science project, Backyard Worlds: Planet 9. The paper is organized as follows: In Section \ref{sec:byw} we describe the citizen science project and the methods of discovery. In Section \ref{sec:align} we discuss the results of our chance alignment probabilities. In Section \ref{sec:obs} we describe the acquisition of spectra and archival data. The results of our analysis are presented in Section \ref{sec:analysis}. Section \ref{sec:benchmarks} discusses the different type of benchmark systems and their distinct uses. We discuss the properties of our sample in Section \ref{sec:context} and place them into context with the literature. Our conclusions are presented in Section \ref{sec:conclude}. We include a discussion of noteworthy and interesting systems in Appendix \ref{sec:noteworthy}.

\section{Backyard Worlds: Planet 9 Citizen Science Candidate Selection}
\label{sec:byw}

\subsection{Selection method 1:  Zooniverse Search} \label{zoo}
Backyard Worlds: Planet 9 (hereafter referred to as BYW) is a citizen science project hosted on the Zooniverse platform with the goal of identifying brown dwarfs missed in previous searches, as well as objects within the Solar System (such as the hypothesised Planet 9) \citep{2017ApJ...841L..19K}. In the BYW project, citizen scientists visually inspect an animated 'flipbook' of time-resolved 704\arcsec\ x 704\arcsec\ coadded unWISE images (for more information see \citealt{2017AJ....154..161M,2017AJ....153...38M,2018AJ....156...69M}). Because of their red {\it WISE} W1--W2 (3.4 $\mu$m -- 4.6 $\mu$m) color and relatively fast proper motions, brown dwarfs and other high proper motion stars stand out in BYW's animated flipbooks. This platform has led to many new discoveries by citizen scientists \citep[e.g.,][]{Faherty21,  meisner2020, Schneider2020, 2021RNAAS...5...76J, 2022RNAAS...6..229G}.

Through their searches, citizen scientists can visually observe that their candidate brown dwarf shares a similar proper motion to another nearby object within the field of view, and flag their submission as a possible co-mover. In addition to just using the default flipbooks presented by BYW, many volunteers also make use of the citizen science developed visualization tool "Wiseview" \citep{2018ascl.soft06004C}, which allows for custom settings (such as FOV, contrast, speed of "flipbook", etc.). In order to create our candidate list, we first collected all user submissions which were flagged as ``comovers''. Each candidate pair was then visually inspected for authenticity. Using the CatWISE2020 catalog \citep{2021ApJS..253....8M}, we gathered the W1 and W2 magnitudes, as well as the proper motions for each candidate brown dwarf (hereafter referred to as the secondary component of the candidate binary system). Proper motions, parallaxes, and photometry of all primary components were taken from the Gaia eDR3 catalog \citep{2021A&A...649A...1G}, which was the latest data release at the time of the analysis. We subsequently used Gaia DR3 values when they became available. Adopting the parallax of the primary component for the secondary, we plotted each secondary on a W1--W2 vs M$_{W2}$ color-magnitude diagram (CMD) as a method to remove any obvious chance alignments.  This was a visual check to eliminate objects that had a $\sim$3$\sigma$ difference between the spectrophotmetric distance of the secondary and the parallax of the primary. For example, an object with the W1--W2 color of a late-type T dwarf but with the absolute W2 magnitude of an M dwarf would be flagged as an unlikely companion candidate. Seventy-four candidate systems resulted from this search.

\subsection{Selection method 2: Gaia to CatWISE Direct Search}
The BYW project has been enormously successful at finding a large number of new brown dwarfs, including co-moving systems.
However, all of the co-moving systems discovered thus-far through BYW have been serendipitous. To supplement the discoveries already made through BYW, we conducted a dedicated search to identify candidate proper motion companions using CatWISE2020 \citep{2021ApJS..253....8M} and Gaia DR2 \citep{2018AandA...616A...1G}, which was the latest Gaia data release at the time of this analysis. All values were updated with those from Gaia DR3 as those became avaialable.

This search was done via a catalog cross match between CatWISE2020 for the brown dwarf secondary components, and Gaia DR2 limited to 100 pc (which included 700,055 sources). CatWISE2020 provides proper motion values for all sources with average astrometric uncertainties on $\mu_{ra}$ and $\mu_{dec}$ of 10 mas yr$^{-1}$ (30 mas yr$^{-1}$) for sources with W1 magnitudes $\sim$12.5 mag ($\sim$15.5 mag; \citealt{2021ApJS..253....8M}). Using a Python script, we queried the CatWISE2020 catalog in a 20 arcminute radius around any Gaia DR2 source with a parallax $>$ 10 mas and required a $<$3$\sigma$ proper motion component match or:

\begin{equation}
    \left| \mu_{\alpha_{Gaia}}-\mu_{\alpha_{Catwise}}\right|<3*\sqrt{\sigma_{\mu_{\alpha_{Gaia}}}^2+\sigma_{\mu_{\alpha_{Catwise}}}^2}
\end{equation}

\begin{equation}
    \left| \mu_{\delta_{Gaia}}-\mu_{\delta_{Catwise}}\right|<3*\sqrt{\sigma_{\mu_{\delta_{Gaia}}}^2+\sigma_{\mu_{\delta_{Catwise}}}^2}
\end{equation}

For sources that passed this criterion, we used the `ab$\_$flags' and `cc$\_$flags' columns in the CatWISE2020 catalog to filter out as many artifacts and false detections as possible. These flags contain a series of letters, corresponding to each of the WISE bands. Each letter represents a possible artifact: `D' for diffraction spike, `H' for scattered-light halo, `O' for optical ghost, or `P' for charge persistence. If the source is believed to be real, but contaminated by an artifact, the same letters are used but in lowercase form. We removed any CatWISE source which contained a capital letter flagging either the W1 or W2 bands. As CatWISE2020 does not contain W3 or W4 photometry, we made no cuts to remove artifacts in those bands. We also did not remove any sources with lowercase letters to avoid cutting any potentially real sources.

What remained afterwards was a list of $\sim$40,000 candidate co-moving pairs to a known Gaia DR2 star. We then applied a color cut so that only sources with a W1--W2 color $>$ 0.2 mag remained, restricting the candidate secondaries to spectral types later than $\sim$M7 \citep{2016A&A...589A..49S}. Next, we estimated the projected physical separation of the system by combining the distance of the primary with the separation between Gaia and CatWISE sources. Based on the separation distribution of known wide co-moving systems with a low mass companion (e.g \citealt{Faherty10}), we eliminated any system with a projected separation $>$ 30,000 AU in order to minimize chance alignments. Since pairs where both components have a detection in Gaia have been the focus of other searches for co-moving pairs (cf.\ \citealt{2021MNRAS.506.2269E,2023RNAAS...7...50Z}), we were only interested in pairs with a CatWISE-only detected companion. We therefore eliminated all pairs in which both objects had entries in Gaia. However, it should be noted that no Gaia-Gaia or $>$30,000 AU pairs were removed from the sample selected in Section \ref{zoo}.

Our remaining list consisted of 5,963 candidate pairs. To verify that each system was authentic (both objects being real, and actual motion seems to match) we only retained pairs which had a total proper motion $\geq$100 mas/yr, resulting in 3,135 candidates. In order to parse through the large number of candidates, we used the machine learning algorithm described in \cite{2019ApJ...881...17M} to help assign a likelihood to each candidate ultracool companion. This machine learning algorithm, based on the python code XGCBoost \citep{10.1145/2939672.2939785}, was trained on red and faint brown dwarfs from the literature, and assigns a “score” on how likely a source in CatWISE2020 is to be real (for further details, see \citealt{2019ApJ...881...17M}). Retaining only sources which had a score $>$90$\%$, the list of candidate pairs was reduced to 1,278 pairs.  Sixty percent of this final list has been visually inspected, and 58 candidates have been identified, with 19 of those overlapping with the candidates identified in Section \ref{zoo}. The remaining sources inspected contained either spurious detections, known co-moving pairs, or companions which did not have visually matching proper motions.

\section{Chance Alignment Probabilities}
\label{sec:align}
In order to assess the probability of each system being a real co-moving pair and not a chance alignment, we used a modified version of BANYAN $\Sigma$, called \texttt{CoMover} \citep{2021ascl.soft06007G}. This code uses the proper motion components, parallax or distance estimate, heliocentric radial velocity (where available), and sky position -- all with uncertainties -- for both the host star and its potential companion. Utilizing a six dimensional multivariate Gaussian in Galactic coordinates and space velocities, a single spatial-kinematic model is constructed from the input of the host star. The observables of the companion are then compared to the host's model, as well as to the field-star model of \cite{2018ApJ...856...23G}. The code also uses Bayes' Theorem to marginalize over any of the companion's missing data with the analytical integral solutions from \cite{2018ApJ...856...23G}. The output of \texttt{CoMover} is a probability that the host and companion are a real, co-moving system.   

\subsection{High Probability Systems}
 Due to the use of Bayes' Theorem in \texttt{CoMover}, the output for a low probability system ($\lessapprox$70$\%$) is inherently unstable. Multiple runs of the code using the same observables can result in different probabilities. However, this has not been observed to produce a false positive, rather the different resultant percentages are all still low. We observed that outputs which are above 90$\%$ appear to converge on the same probability. We therefore make a conservative cutoff at 90\% as what constitutes a ``high probability'' co-moving system.
 
 Seventy-nine percent of the candidate pairs identified in this paper received a co-moving probability above this threshold, resulting in 89 high probability systems. A summary of these high likelihood systems is listed in Table \ref{table: 3}, with their respective photometry shown in Table \ref{table: 2}. Hosts and companions are shown on CMDs (using the host star's parallax for the companions) in Figure \ref{fig:1}. 

\subsection{Low Probability Systems}\label{low-p}

Twenty-four of the candidate pairs considered received a co-moving probability below our 90$\%$ threshold using \texttt{CoMover}. These twenty-four systems, which were originally identified as potential co-moving pairs, thus have a high likelihood of being chance alignments, and we therefore do not consider them in the discussion which follows. Three of the ultracool dwarfs in these low likelihood pairs had been followed up spectroscopically, and are presented in Figure \ref{fig:12}. A summary of the low likelihood pairs is shown in Table \ref{table: 10}.

\section{Data Acquisition}
\label{sec:obs}
\subsection{Spectroscopic Observations}
We have obtained follow-up spectroscopy of 34 of the potential companions identified in Section \ref{sec:byw}. Thirty-three candidate hosts were also observed. A summary of the observation dates and instruments used is shown in Table \ref{table: 1}.

\subsubsection{IRTF/SpeX}
We observed twenty-two objects with NASA's Infrared Telescope Facility (IRTF) located on MaunaKea, Hawaii. Using the SpeX instrument \citep{2003PASP..115..362R} we obtained near infrared spectra over the 0.8--2.5 $\mu$m wavelength range. All objects were observed with SpeX in prism mode with a 0\farcs8 wide slit, providing a resolving power of $\approx$100-500. Additionally, one target was observed using SpeX in the short-wavelength cross-dispersed mode (SXD), with a wavelength coverage of 0.8--2.42 $\mu$m with a resolving power of R$\sim$2000. After each target, a standard A0 star was observed for telluric correction and flux calibration. The data was reduced with Spextool \citep{2004PASP..116..362C} using the telluric and flux calibration techniques of \cite{2003PASP..115..389V}.

\subsubsection{Keck/NIRES}
Ten objects were observed using the Near Infrared Echellette Spectrometer (NIRES; \citealt{2004SPIE.5492.1295W}) at the W. M. Keck II telescope, located on Maunakea, Hawaii. Near infrared spectra (0.94--2.45 $\mu$m) were obtained at a resolving power of $\approx$2700. 
Data were acquired during five nights on 27 October 2018, 14 February 2019, 24 February 2021, 21 July 2021, and 11 June 2022 (UT). Conditions were generally clear with 0$\farcs$5 to 1$\farcs$0 seeing, with the exception of clouds on 27 October 2018 and 21 July 2021. Each source was acquired with the K-band imaging channel of NIRES and observed with the 0$\farcs$55 $\times$ 18$\arcsec$ slit aligned to parallactic, and 4-8 exposures were acquired in an ABBA 10$\arcsec$ dither sequence with integration times of 10~s to 300~s depending on source brightness and conditions. We observed an A0~V star before or after each science target for telluric absorption and flux calibration, and dome-reflected flat-field and arc lamps for order tracing, pixel calibration, and initial wavelength calibration. All data were reduced using a modified version of SpeXtool \citep{2004PASP..116..362C,2003PASP..115..389V} following standard reduction procedures. 

\subsubsection{Lick/Kast}
Twenty objects were observed using the Kast double spectrograph \citep{kastspectrograph} on the Shane 3m Telescope at the Lick Observatory located on Mount Hamilton, California. Data were acquired over six nights on 17 May 2021, 12 December 2021, 29 September 2022, 31 October 2022, 1 June 2023, and 12 August 2023 (UT) in conditions that ranged from clear to thin cirrus and seeing of 1$\farcs$1--1$\farcs$6. All sources were observed using the 600/7500 grism and 1$\farcs$0- or 1$\farcs$5-wide slit, providing 6000--9000~{\AA} wavelength coverage at an average resolution of $\lambda/\Delta\lambda$ = 1900--2800. Individual exposures ranged from 30~s to 1200~s depending on source brightness, and were obtained in pairs to reduce cosmic ray contamination. For each source, we observed a nearby G2~V star for telluric absorption correction, and on each night we observed a flux calibrator drawn from \citet{1992PASP..104..533H,1994PASP..106..566H}. Data were reduced using the kastredux package\footnote{kastredux:~\url{https://github.com/aburgasser/kastredux}.} using the baseline parameters for batch reduction. 

\subsubsection{SOAR/TripleSpec4}
Three objects were observed with the SOAR Telescope at the Cerro Tololo Inter-American Observatory (CTIO), using the Astronomy Research using the Cornell Infra Red Imaging Spectrograph (ArcoIRIS; \citealt{schlawin2014}) otherwise known as TripleSpec4. Near infrared spectra (0.8--2.4 $\mu$m) were obtained across 6 simultaneously observed
cross-dispersed orders covering the 0.8 - 2.4 $\mu$m range,
with a resolving power of $\approx$3500. Science exposures
were taken at two different nod positions along the slit,
which has a fixed width of 1.1$\arcsec$. After each science target,
A0V stars were observed in order to execute telluric corrections. Data were obtained on 2021 October 23 under stable and clear conditions.  Reductions were performed using a modified
version of the SpeXtool reduction package (\citealt{2004PASP..116..362C,2003PASP..115..389V}).

\subsubsection{SALT/RSS}
Seven objects were observed with the Robert Stobie Spectrograph (RSS)\citep{2003SPIE.4841.1463B, 2003SPIE.4841.1634K}, located on the Southern African Large Telescope (SALT)\citep{2006SPIE.6269E..0AB}. The long slit mode of the spectrograph was used, with the PG0900 grating at a 20$^o$ angle. This gives coverage over 3 wavelength ranges: 6033-7028 \AA, 7079-8045 \AA, and 8091-9023 \AA. Neon arc lines were observed immediately after each target, which was used to wavelength calibrate the observatory-provided pre-processed data (included gain and cross talk correction, as well as overscan subtraction). The target's spectrum was then flux calibrated using the \cite{1994PASP..106..566H} standard EG21, which was acquired on 2023 January 24 UT using the same setup.

\subsection{Catalog Photometry, Astrometry, and Abundances}
 We performed a literature search for any existing near-infrared photometry in the Two Micron All Sky Survey (2MASS) \citep{2006AJ....131.1163S}, various UKIRT surveys (e.g., UKIDSS; \citealt{2007MNRAS.379.1599L}, UHS; \citealt{2018MNRAS.473.5113D}), and from the VISTA science archive \citep{2012A&A...548A.119C} for both host and companion objects. 

For all candidate host stars we gathered proper motions, parallaxes, radial velocities, and rotational broadening ($v$sin$i$) using Gaia DR3 \citep{gaia2023}. Additionally, Gaia DR3 contains abundances and metallicity for select sources, which were also gathered if present. Further literature searches were performed for any previously reported abundances or metallicity. For sources which had metallicity measurements available in the literature, no corrections were added to account for systematic offsets between different studies.   

\subsection{TESS Light Curves}
As the characterization of brown dwarfs is heavily age dependent, being able to derive an estimate for the host star age in a benchmark system is crucial. Gyrochronology, the use of a star's rotation rate to constrain its age \citep[e.g.,][]{2003ApJ...586..464B,2007ApJ...669.1167B}, has been shown to provide reliable age constraints for solar and later type stars for ages of a few hundred Myr out to field age. With new large area variability surveys, such as NASA's Transiting Exoplanet Survey Satellite (TESS; \citealt{2015JATIS...1a4003R}), obtaining rotation rates for large numbers of stars has become much easier, increasing the power and usefulness of gyrochronology as an age diagnostic tool.

Host sources were checked to see if any had been observed by TESS with either 30 minute or 2 minute cadences. If observed, a light curve was generated using simple aperture photometry and background subtraction. Each light curve was then run through a Lomb\-Scargle periodogram \citep{1976Ap&SS..39..447L,1982ApJ...263..835S}, resulting in a best fit rotation period (see \citealt{Popinchalk23} for further details regarding the process used).

\section{Analysis}
\label{sec:analysis}
\subsection{Spectral types}
\subsubsection{Primary Components}
\begin{figure*}
    \centering
    \includegraphics[scale=1]{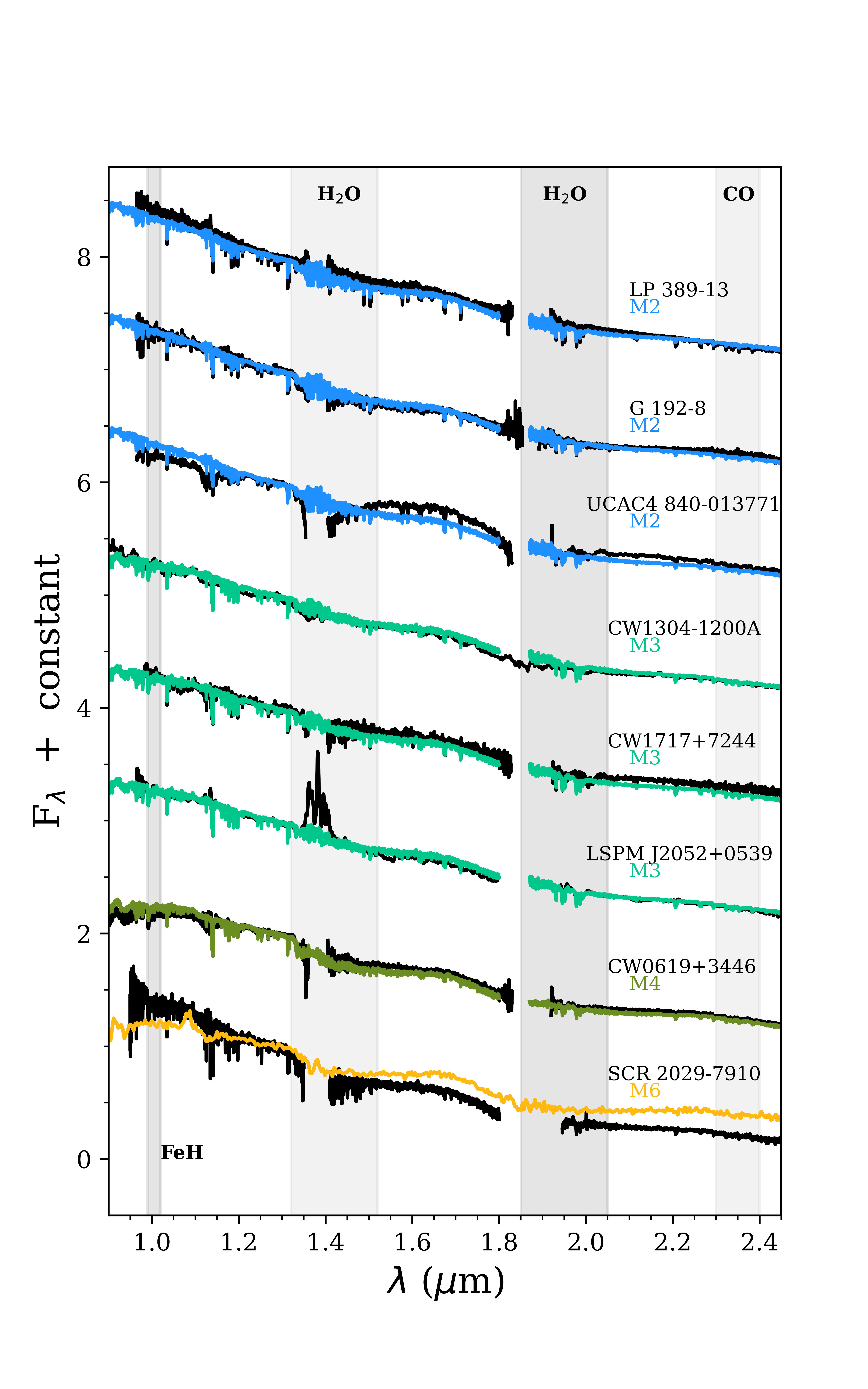}
    \caption{Near-Infrared spectra of the primary components plotted in black. Over-plotted in color are the spectral standards. All spectra were normalized between 1.27-1.29 $\mu$m and separated by a constant. Spectra obtained with: NIRES-- G 192-8, CW1717+7749, LP 389-13, LSPM J2052+0539; SpeX-- CW0619+3446, CW1304-1200; ArcoIRIS: SCR 2029-7910.}
    \label{fig: 5}
\end{figure*}

For candidate primaries which were observed spectroscopically, spectral types were assigned by overplotting either infrared spectral standards (using the standards of \citet{Kirkpatrick2010} for types M7 and later, and sources from the IRTF/SpeX spectral library for sources earlier than M7) or optical standards (using standards from \cite{1991ApJS...77..417K} for spectral types K5 through M9, and \cite{2017ApJS..230...16K} for earlier types) on top of the observed data, normalizing all spectra and standards to the same wavelength. In the optical, spectral types were assigned via a visual match to the entire wavelength coverage, whereas spectral types in the near-infrared were assigned based on a visual match of the J band portion of the spectrum following the technique of \citet{Kirkpatrick2010}. The primary near-infrared and optical spectra are shown in Figures \ref{fig: 5} and \ref{fig: 6} respectively.

For candidate primaries which did not have spectral types or phototypes in the literature, and were not observed spectroscopically, we used the Gaia parallax and photometry along with the absolute magnitude relations of \cite{2019AJ....157..231K} to estimate spectral types. If the candidate primary lacked a Gaia parallax, its spectral type was estimated from the spectral type versus color relations from \cite{2021ApJS..253....7K}.

\subsubsection{Secondary Components}
\begin{figure*}
    \centering
    \includegraphics[width=1\textwidth]{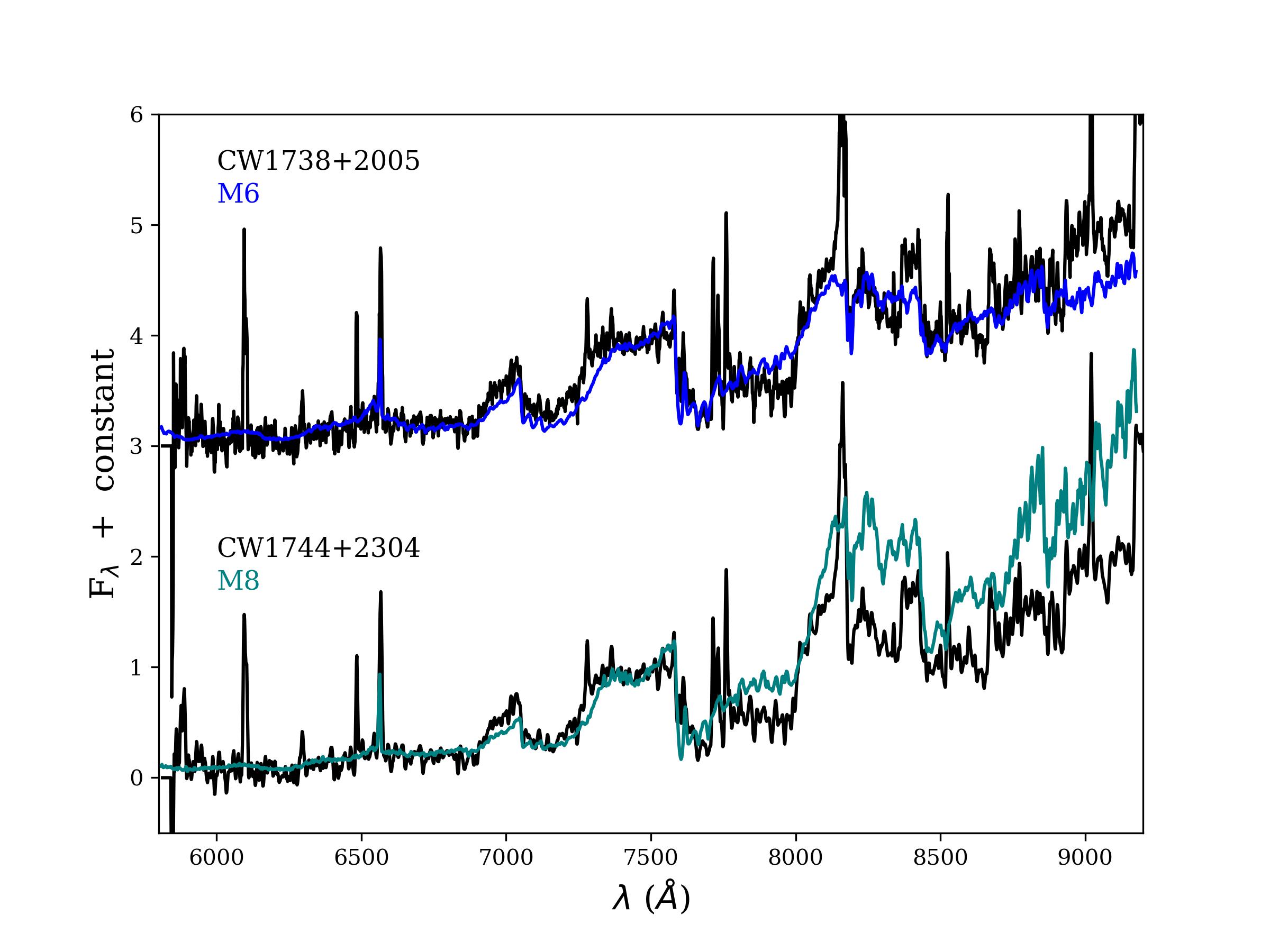}
    \caption{Optical spectra from Lick/Kast of the secondary components plotted in black. Over-plotted in color are the spectral standards. All spectra were normalized between 7499-7501 \AA\ and separated by a constant.}
    \label{fig: 4}
\end{figure*}

 For the secondaries, spectral types were assigned by overplotting the infrared spectral standards from \cite{2006ApJ...637.1067B} and \citet{Kirkpatrick2010} on top of the observed spectra and visually comparing the peak of the J band. All secondary spectra are shown in Figures \ref{fig: 3} (near-infrared spectra) and \ref{fig: 4} (optical spectra).

For candidate secondaries with no available spectra, we estimated the spectral type by adopting the host star's parallax and using the spectral type versus $M_{W2}$ relation from \cite{2021ApJS..253....7K}. If the candidate secondary was a companion to an object which lacked a parallax, the secondary's spectral type was estimated with the spectral type versus color relations from \cite{2021ApJS..253....7K} For candidate secondaries which had their own Gaia astrometry and lacked spectra, spectral types were estimated using the photometric relations from \cite{2019AJ....157..231K}.

\subsection{Low Probability Systems}
While the low probability systems presented in Section \ref{low-p} have a high likelihood of being chance alignments, due to large projected physical separations and differences in proper motions, 6 systems still receive comoving probabilities $>$60$\%$. Of these, 3 systems have probabilities $\>$85$\%$: G 1-44 + 2MASS J0102+0355 (88.6$\%$), WDJ1324-1902 + CW1325-1907 (87.5$\%$), and 2MASS J2308-5052 + CW2308-5052 (87.2$\%$). These systems, while not above the ``high probability'' threshold, still have elevated probabilities of being associated with one another. These candidate companions have proper motion errors which range from $\sim$34-70\arcsec, making it difficult to confirm their comoving nature. A re-analysis with higher accuracy proper motions is required in order to further investigate the candidacy of these systems. 

\begin{figure*}
    \centering
    \includegraphics[width=1\textwidth]{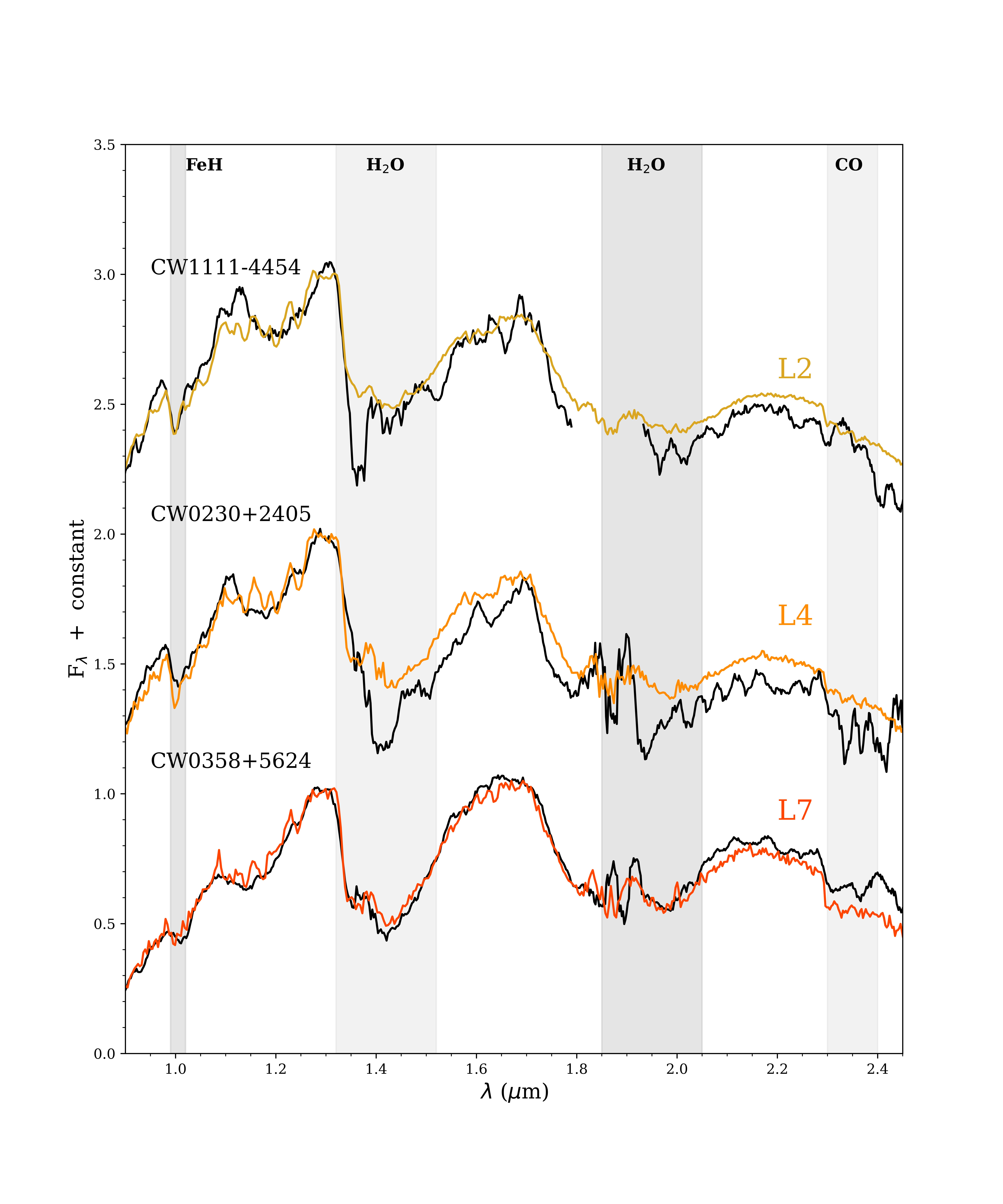}
    \caption{Spectra obtained with SpeX for low probability companions plotted in black. Over-plotted in color (with redder colors indicating cooler spectral temperatures) are the spectral standards as described in Section \ref{sec:analysis}. All spectra have been normalized between 1.27-1.29 $\mu$m and separated by a constant.}
    \label{fig:12}
\end{figure*}

Three of the ultracool dwarfs in the low probability sample were followed up spectroscopically: 

\textbf{CW1111-4454} Comparing the near-IR spectrum of CW1111-4454 to that of the spectral standard in Figure \ref{fig:12}, we place CW1111-4454 as spectral type L2 (sl. blue), due to the slightly supressed K-band of CW1111-4454 which gives it a slightly bluer spectral morphology. 

\textbf{CW0230+2405} We place CW0230+2405 as spectral type L4 (blue) through comparison with the L4 spectral standard as shown in Figure \ref{fig:12}. Both the H- and K- bands of CW0230-2405 appear to be suppressed compared to the standard. 

\textbf{CW0358+5624} CW0358+5624 is best fit as spectral type L7, as shown in Figure \ref{fig:12}.

\subsection{Host RUWE}
 The Gaia re-normalized Unit Weight Error (RUWE) \citep{2018A&A...616A...2L} is related to how `well behaved' the astrometric and photometric solution is for a source when compared to the proper motion and parallactic motion expected for a single source, while taking into account the source's magnitude and color. A `perfectly well behaved' RUWE value would be expected to be $\sim$1.0. Deviations from this value are caused by non-linear motions of the photocenter of the source in the Gaia images, as well as unusual colors (e.g. as a product of variability). Therefore, an elevated RUWE value can be attributed to any potential variability of the source, as well as motion caused by a binary orbit. 

Several studies have been performed on testing the RUWE with known binary systems, and it has been shown that the RUWE strongly correlates with unresolved binary companions \citep{2021ApJ...907L..33S, 2020MNRAS.496.1922B}. Nominally, RUWE values $\gtrsim$1.4 are considered to be poorly behaved, although \cite{2021ApJ...907L..33S} have shown that even RUWE values just slightly larger than 1.0 can be indicative of an unresolved companion. 

We gathered the RUWE values for all host stars, and find that 20 sources have a RUWE $\>$1.4, as shown in Table \ref{table: 8}. 

\subsection{Gravity Indices}
Young ultracool dwarfs, which have not yet fully contracted, should have lower surface gravities than their field counterparts. Thus, being able to determine the surface gravity of an ultracool dwarf can provide insight into the object's age. 

For objects between M6 - L7 which had near infrared spectra, we calculated the FeH-z, VO-z, KI-J, and H-cont indices of \cite{2013ApJ...772...79A}, which have been found to be sensitive to surface gravity. These indices are not appropriate for spectral types earlier than M6 or later than L7, so these spectra were left out of the analysis. Following the approach of \cite{2013ApJ...772...79A}, we score each of the indices based on spectral type, and assign an overall classification for the object of field gravity (FLD-G), intermediate gravity (INT-G), and very low gravity (VL-G). We find 7 companions which receive an INT-G classification, and 19 which are classified as FLD-G. These results are shown in Table \ref{table: 4}.

\subsection{Spectral Binary Indices}
\label{sec:sb}
For ultracool dwarfs in tight binary pairs which are unresolved, the combined light from the two objects can result in an observed spectrum resembling that of a single, typical ultracool dwarf \citep{2010ApJ...710.1142B,2014ApJ...794..143B}. These systems are referred to as spectral binaries. If the two components of the spectral binary are different enough from one another (such as an L- and T-dwarf), the observed spectrum is likely to contain peculiarities which arise from the different spectral features between the two objects. 

In order to investigate if any of the observed spectra of the candidate companions in our sample are indicative of spectral binarity, we followed the methods of \cite{2014ApJ...794..143B} for spectral types M7--L7 and \cite{2010ApJ...710.1142B} for spectral types L8--T5. Table~\ref{tab:sb} lists the indices calculated for each object, including the strength of spectral binary candidacy based on the definitions of \cite{2010ApJ...710.1142B} or \cite{2014ApJ...794..143B} (weak,  strong, or rejected). Of the 29 companions included in this analysis, 2 objects were found to be strong spectral binary candidates, 7 objects were found to be weak spectral binary candidates, and 20 objects were rejected as spectral binaries. Futher discussion of these SB candidates can be found in Section \ref{sec:sb_diss}. 

\subsection{Rotation Periods}

\begin{figure*}
    \centering
    \includegraphics[width=1\textwidth]{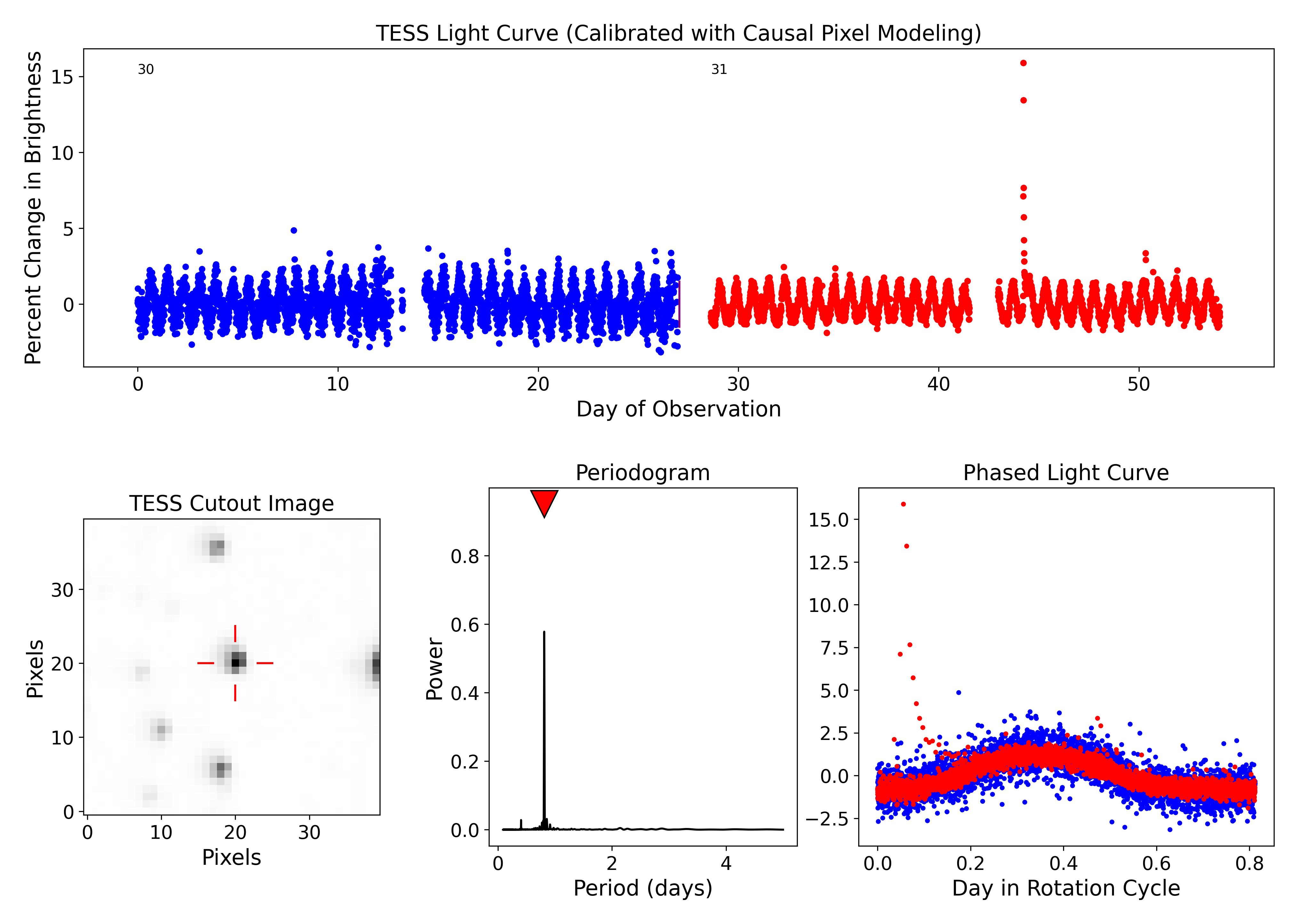}
    \caption{\textbf{Top}: TESS extracted light curve of CW0256-3350 from TESS sectors 30 (shown in blue) and 31 (shown in red). A flare can bee seen in TESS sector 31 at day $\sim$45. \textbf{Bottom-Left}: Cutout image of the TESS field of view around CW0256-3350. \textbf{Bottom-center}: Lomb\-Scargle periodogram for the light curve of CW0256-3350, showing a rotation period of 0.81 days. \textbf{Bottom-right}: Phase-folded light curve for CW0256-3350.}
    \label{fig: 2}
\end{figure*}

\begin{figure*}
    \centering
    \includegraphics[width=1\textwidth]{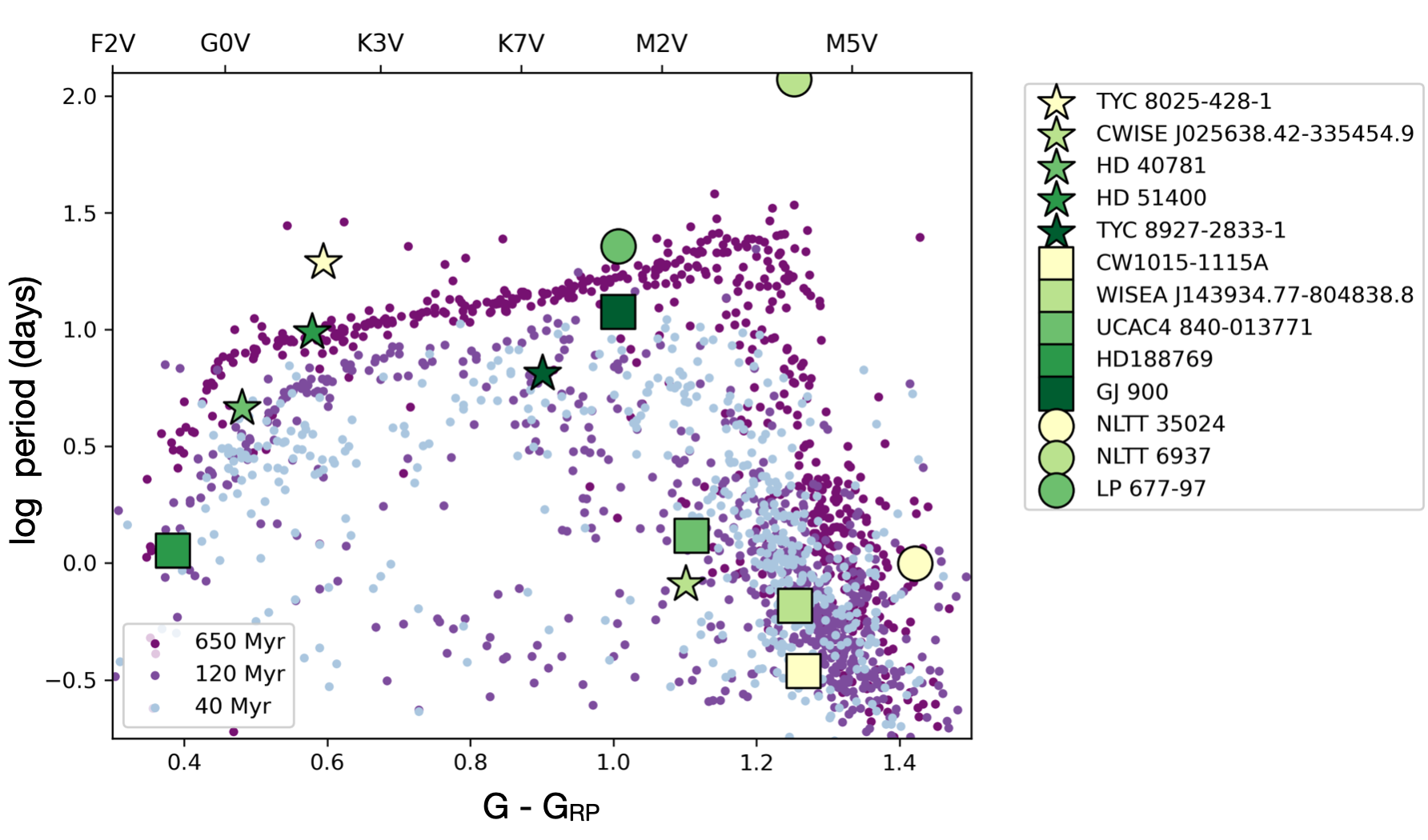}
    \caption{Gaia color vs. logarithm of rotation period for host stars with observed light curves (Pointed stars, squares, and circles). For age comparison, host stars are plotted against the 650 Myr Praesepe \citep{2017ApJ...842...83D} (in dark purple), 120 Mry Pleiades \citep{2016AJ....152..113R} (lighter purple), and 40 Myr Tucana-Horologium \citep{2023ApJ...945..114P} (lightest purple) groups.}
    \label{fig: 9}
\end{figure*}

Seventy-three of the host stars in our sample were bright enough to be detected with TESS. Of these seventy-three, we were able to confidently measure a rotation period for $\sim$14$\%$ (10 sources). These sources are listed in Table \ref{table: 5}. A sample light curve and periodigram from our workflow is shown in Figure \ref{fig: 2}. We placed these stars on a color-period diagram along with stars from the Tucana-Horologium (40 Myr, \citealt{2023ApJ...945..114P}), Pleiades (120 Myr, \citealt{2016AJ....152..113R}) and Praesepe (650 Myr, \citealt{2017ApJ...842...83D}) clusters as age benchmarks, shown in Figure \ref{fig: 9}.

\subsection{BANYAN $\Sigma$}
\label{sec:ban}
Each candidate system was checked for young moving group memberships probabilities using BANYAN $\Sigma$ \citep{2018ApJ...856...23G}. BANYAN $\Sigma$ is a Bayesian algorithm which uses the spatial and velocity coordinates of a source in order to identify potential membership to young moving groups (YMG). For each source, BANYAN $\Sigma$ uses its sky coordinates, proper motions, radial velocity, parallax, and uncertainties, comparing them in full 6D configuration space (XYZUVW) to 27 known young associations (see \citealt{2018ApJ...856...23G} for a detailed explanation). BANYAN $\Sigma$ also allows for the possibility of using only sky coordinates and proper motions if no other measurements exist. 

The output of BANYAN $\Sigma$ is a list of probabilities for membership to each of the 27 YMGs, along with a probability of belonging to the "field" population. The ages of the YMGs considered in BANYAN $\Sigma$ range from $\sim$1-800 Myr. As BANYAN $\Sigma$ only contains 27 YMGs, a source receiving a high probability of belonging to the field can only be said to not likely belong to any of those 27 groups. There still exists the possibility the source may still belong to one of the other known YMGs which are not included in the analysis of BANYAN $\Sigma$.

 For each candidate system in our sample, we ran all available kinematics for both host and companion through BANYAN $\Sigma$ in order to check for any potential YMG memberships. Systems which received a likely membership are discussed further in Appendix \ref{sec:noteworthy}.
   
 \section{Discussion of Benchmark Systems}
 \label{sec:benchmarks}

Benchmark systems are extremely useful, as they provide a means of gaining insight on fundamental properties like distances, chemical abundances, and age from a co-moving system where one object is very well characterized and can inform on the other. The sample of ultracool benchmark systems presented in this paper comprise a wide variety of host spectral types, each of them being able to benchmark different aspects of their companion's properties. Below we describe the different ``populations'' of brown dwarf benchmarks in our sample, as well as the vital role each can play.

\subsection{Age Benchmarks}
Twenty-four of the host stars presented in this sample provide a means of reliably constraining an age range for their ultracool dwarf companions. These ages are listed in Table \ref{table: 6}. This added information makes it possible to break the mass-temperature-age degeneracy of substellar mass objects, which will be utilized in future works aiming to derive fundamental properties of the BYW benchmark sample. 

Six of the hosts (or at least one component of the host system) in our sample are white dwarfs: WDJ060159.98-462534.40, LP 369-22, UCAC4 328-061594, WDJ002027.83-153546.50, WDJ011001.84-592642.22, and WDJ212231.86+660043.69. Using the python code \texttt{wdwarfdate} \citep{2022AJ....164...62K}, we input the effective temperature and log(g) of the white dwarf from \cite{2021MNRAS.508.3877G}, and get the Bayesian estimated white dwarf mass, progenitor star mass, white dwarf cooling age, and the total system age (including the estimated lifetime of the progenitor star). The ages obtained for these sources range from $\sim$0.42-10.91 Gyr. \cite{2022AJ....164...62K} state that the cooling tracks employed by \texttt{wdwarfdate} assume a C/O core, and therefore the ideal mass range for this method is between 0.45-1.1 $M_{\sun}$. Likewise, \cite{2022AJ....164...62K} point out that white dwarfs with masses $<$0.45 $M_{\sun}$ and $>$1.1 $M_{\sun}$ may have potentially formed as a result of binary mergers/evolution. Results for white dwarfs outside of this ideal range should be treated with extreme caution. Two of our white dwarf hosts lie outside of this range: WDJ011001.84-592642.22 (M = 0.43 $M_{\sun}$) and UCAC4 328-061594 (M=1.19 $M_{\sun}$); these sources require further study to more accurately constrain their ages. For more information on the limitations and errors of \texttt{wdwarfdate}, refer to \cite{2022AJ....164...62K}. 

While we do believe a seventh host star to be a white dwarf, its photometry is contaminated, preventing us from properly estimating its age. Further observations of this source, however, should provide enough information to be able to constrain its age.

Nine of our hosts are main sequence stars with age estimates in the literature, spanning $\sim$0.1-4.2 Gyr. Seven of these age estimates utilize the equivalent width of the CaII infrared triplet excess from \cite{2017ApJ...835...61Z}, noting that this approach only allows us to place loose age constraints on these systems (i.e. 0.1--1 Gyr or $>$1 Gyr). 

For ten of our host stars, we were able to obtain a light curve from TESS observations. An additional 3 of our host stars have reported rotational periods in the literature which are not derived from TESS. We compare these sources to confirmed members of the Pleiades (120 Myr) and Praesepe (650 Myr) groups on a color-period diagram (see Figure \ref{fig: 9}). From comparing the rotation periods of our objects with that of these benchmark clusters, we estimate that all but two sources appear to be $<$1 Gyr, with 5 appearing to be potentially younger than the $\sim$650 Myr Praesepe cluster, based on their rotation periods. These ages should be considered estimates based on comparison to benchmark populations and not derived from robust gyrochrones. Two of these hosts stars have a companion which received an INT-G gravity score, while the companions for the rest either lack spectra or lack clear gravity sensitive indices for their spectral type. 

One of our host stars, GJ 900, is a potential member of the nearby moving group Carina-Near, which has an age of 200$\pm$50 Myr \citep{2006ApJ...649L.115Z}. 

\subsection{Compositional Benchmarks}
One of the most difficult properties to measure for an ultracool dwarf is chemical composition (e.g. metallicity, C/O ratio, Mg/Si ratio). Chemistry is challenging to derive for most stars given that high resolution spectroscopy is required and solar type stars are the optimal spectral type for measurements. The intrinsic faintness of ultracool dwarfs makes obtaining the high-resolution spectra necessary for these measurements difficult for many objects. To date, only $\sim$50 wide, resolved ultracool companions with a spectral type of L or later to solar-type analogs (FGK stars) are known \citep[Calamari et al. in prep]{2010AJ....139..176F, 2014ApJ...792..119D}. In this paper, we present an additional 21 ultracool companions to FGK stars, consisting of 3 late M dwarfs, 14 L dwarfs, and 4 T dwarfs, expanding the known widely separated companions to FGK stars by $\sim$42\%. Each of the host stars has the potential to be chemically characterized and then the secondary can have its observational features calibrated.  Ten of the host systems have available abundances in the literature thanks to large area surveys such as LAMOST or Gaia DR3. These systems, and their published abundances are listed in Table \ref{table: 7}. 

There are two ways the ultracool dwarf compositional benchmarks can be utilized:  (1) with a large enough sample of L and T dwarfs co-moving with an FGK star, we can correlate the host star compositions with observational properties of the secondary.  For instance, we can look for metallicity sensitive features across the near infrared and map that by the metallicity measurements of the primary.  (2) Utilizing the well defined properties of the host star, we can constrain the retrieved properties of the companion (such as metallicity and abundances), testing ultracool dwarf and exoplanet atmospheric models. 
 
Fifty-five of our host stars have metallicities reported in the literature, shown in Table \ref{table: 7}. The range of metallicites (both [Fe/H] and [M/H]) for these stars are from -2.03 to 0.55 dex, with fifteen having a metallicity $<$ -0.5 dex. Adopting the host's metallicity for their ultracool companions, the wide range of metallicities in this sample can help address the role it plays in shaping the spectral energy distributions of cold atmospheres. 

Two of the hosts in our BYW benchmark sample allow us to derive C/O abundance ratios from their reported [C/H] and [O/H] metallicites. The range of these abundances are from C/O=0.44 to 0.79. While potentially useful in studying ultracool dwarf formation pathways, comparing the C/O ratios between a host and its ultracool companion can serve another important purpose. \cite{2022ApJ...940..164C} point out a trend in the retrieved C/O values in late-T dwarfs to be supersolar. While this could be hinting at formation, \cite{2022ApJ...940..164C} postulate that this trend is likely due to unconstrained CO and CO$_2$ abundances, as well as potentially misunderstood or unconstrained condensate chemistry. Many of these late-T dwarfs with retrieved C/O values do not have a companion with which to compare their abundances, however. While the retrieval work is deferred to future papers, the sample presented in this paper allows for more direct comparisons between the atmospheric abundances retrieved in ultracool dwarfs with their host, hopefully beginning to shed more light on the observed trends.

Three of the hosts in our BYW benchmark sample allow us to derive Mg/Si abundance ratios from their reported [Mg/H] and [Si/H] metallicities. The Mg/Si abundances for these sources ranges between Mg/Si=0.68 to 1.4. In many ultracool atmospheres, the temperatures become cool enough for the formation of silicate and iron clouds \citep{2006asup.book....1L}. From models of brown dwarf atmospheres, these clouds become especially prominent in the mid-late L dwarfs and across the L-T transition, until they sink below the photosphere in T-dwarfs \citep{2006asup.book....1L}. Observations agree nicely with this general picture, with dust clouds being found to peak between the L4 and L6 spectral types, sedimenting at spectral type $\sim$ L8 \citep{2022MNRAS.513.5701S, 2023MNRAS.523.4739S} The species of silicate clouds which can form in these atmospheres has direct implications in modelling the observed atmosphere, as different silicate clouds will sequester different amounts of oxygen as well as shape the temperature - pressure profile. Atmospheres with a high Mg/Si (magnesium rich) are more likely to form forsterite (Mg$_2$SiO$_4$) and enstatite (MgSiO$_3$) clouds, however for low Mg/Si atmospheres (silicon rich), enstatite and quartz (SiO$_2$) are the expected species (Calamari et al. in prep). Assuming both host and companion formed with the same elemental abundances, knowing the Mg/Si ratio of the host allows for a prediction of the clouds expected in the companion, directly testing the retrieved cloud species. With a large enough sample of ultracool dwarf companions which can have their atmospheric chemistry anchored via their host's composition, it may be possible to begin constraining theories of cloud formation and sedimentation in both ultracool dwarfs and exoplanets. The host stars in this sample provide a large range of Mg/Si abundances,  which can be utilized to directly test varying cloud formation models in their ultracool companions.

\section{BYW Benchmark Sample in Context with Literature Benchmark Sample}
\label{sec:context}
Within this section we summarize several critical properties of the new co-movers in the context of the literature sample of benchmarks. 

\subsection{Distance distribution of The BYW Benchmark sample}
 The overall distances covered by the full BYW benchmark sample ranges from $\sim$20 pc out to $\sim$170 pc, with the peak of the distribution around 40 to 60 pc. For the ultracool companions which are classified as late M's, the distance distribution is from 45 pc to 174 pc. While the majority of these companions, which were discovered using the method described in Section \ref{zoo}, are also detected by Gaia along with their hosts, four companions have no Gaia counterpart: CW1240+4605, CW1325+0223, CW0159+1055, CW1031+1237. The distances for these objects (adopting the Gaia parallax of their host) are 124.0$_{-4.8}^{+5.2}$ pc, 119.0$_{-0.4}^{+0.4}$ pc, 111.0$_{-0.6}^{+0.6}$ pc, and 99.0$_{-0.2}^{+0.2}$ pc respectively. The Gaia Catalog of Nearby Stars (GCNS, \citealt{2021A&A...649A...6G}) is 95$\%$ complete out to 100 pc down to spectral type M8, so it is not unexpected for objects with spectral type M9 beyond 100 pc to be missing astrometry. Additionally, two companions, both spectral type M9, have a Gaia DR3 catalog entry with no astrometry indicating that they were likely at the edge of detection for the mission. 

The distance of the L-dwarf type companions in the BYW benchmark sample ranges from 30 pc to 146 pc, with the distribution peaking around $\sim$55-60 pc. Two of the companions which had been identified using the method described in Section \ref{zoo} have entries in Gaia DR3: CW2003$-$1422 (spectral type L0 at 56.99 pc) and CW2029$-$7910 (spectral type L1 (blue) at 50.95 pc). One companion, CW1303+5127 (spectral type L2), has a Gaia DR3 catalog entry with no astrometry. 

The distances of the T-dwarf type companions in the BYW benchmark sample ranges from 17 pc to 57 pc, with the distribution peaking around $\sim$50 pc.

\subsection{Separation of BYW Benchmark Systems}
Characterizing the demographics of ultracool companions is important for testing and constraining different formation mechanisms \citep{metchev_palomarkeck_2009, 2014ApJ...784...65B, brandt_statistical_2014}. While the occurrence rates and characteristics of widely separated ($>$ 100 AU) ultracool companions have been well studied, only $\sim$46 ultracool companions have been discovered with separations $\geq$1000 AU (see \citealt {2020A&A...633A.152C}). In this paper, we add an additional seventy-two systems to the list of systems with separations $\geq$1000 AU, almost tripling the current census. 

Presently, some of the widest known confirmed ultracool companions have been observed to have projected separations $\sim$15,000 AU \citep{2010MNRAS.404.1817Z,2011MNRAS.410..705D}. The BYW benchmark sample includes twelve systems with companions at separations $>$15,000 AU. Interestingly, seven out of the twelve of these widest systems have T dwarf companions. The formation of these systems is difficult to explain with current models, as \textit{in situ} formation of binary systems via direct collapse and fragmentation of the molecular cloud seems to only form pairs out to $\sim$10,000 AU \citep{2013ApJ...768..110C}. One possible explanation for their formation is that these wide systems are actually of higher order multiplicity, where the interactions between components result in an extremely wide orbit \citep{reipurth_formation_2012, 2016MNRAS.459.4499E}. 

Indeed, we find that five of our wide companions have hosts with multiple components: CW0915+2547 (estimated T5) with K6 and DZ white dwarf hosts; CW1959-6443 (estimated T6) with F3IV and M5 dwarf hosts; SDSS 1344+0839 (T0, identified as a low probability L5+T2 binary) with an M5.5 host which is likely binary (Gaia RUWE=8.3);  CW1625+7749 (L4) with candidate binary (RUWE=1.6) M2 dwarf host, and CW0407+1909 (estimated T1) with M4 and M4 dwarf hosts. Additionally, two of the companions, CW1959-6443 and CW0407+1909, appear overluminous on the WISE CMD, indicating the potential for hidden binarity in these components as well.  

Three of the systems with separations $>$15,000 AU have hosts (or a host component) which are white dwarfs: CW0915+2547, CW1010-2423, and CW2123+6556. As the progenitor star undergoes its transformation into a white dwarf, the mass loss taking place may cause the orbital separation of the host and companion to increase. \cite{2002MNRAS.331L..41B} showed that this increase should scale as $M_{ms}$/$M_{wd}$, where $M_{ms}$ is the initial mass of the progenitor star, and $M_{wd}$ is the final mass of the white dwarf. Therefore if a system formed with a separation of 5,000 AU, and the progenitor star was five times as massive as its white dwarf stage, the companion would eventually migrate out to 25,000 AU. While this migration may in fact lead to the pair dissolving as the binding energy becomes too small, because of the lack of a sudden and violent interaction, the pair could retain a similar proper motion and distance for a significant time (before drifting apart and being thermalised by the disk population). 

The widest of the systems in the BYW benchmark sample, CW2106+2507, is an estimated T1 dwarf which was given a 100\% probability of co-moving with a K6. The projected separation of the pair at the Gaia distance of the host is $\sim$38,500 AU. Neither component shows obvious signs of hidden binarity, making it unlikely this pair is still gravitationally bound to one another, as their binding energy would be too small to survive the gravitational interactions with surrounding objects. Their receiving a high probability of co-moving with one another hints at them still being associated with one another in some way, however, and could potentially be indicative of a dissolving moving group. 

\subsection{Distribution of Host Mass} \label{sec:mass_gather}
Host mass estimates were taken from \cite{2018AJ....156..102S} and \cite{2019AJ....158..138S} for main sequence hosts, and \cite{2021MNRAS.508.3877G} for WD hosts. When no published mass was present, we estimated its mass using the relations of \cite{2013ApJS..208....9P}.

The host stars in the BYW benchmark sample cover a wide range of masses, from $\sim$0.1 $M_{\sun}$ out to $\sim$1.4 $M_{\sun}$. The peak of the distribution is around 0.2-0.3 $M_{\sun}$, or spectral types between $\sim$M3-M6. Previous studies have found no discernible difference in the brown dwarf companion fraction based on host spectral type (\citealt{2015ApJS..216....7B} and references within), so the observed abundance of mid-M dwarf hosts in our sample is likely not due to a preference for these spectral types to host ultracool companions, but is instead due to these spectral types representing the peak of the spectral type distribution, with mid-M dwarfs comprising $\sim$75$\%$ of the solar neighborhood \citep{2006AJ....132.2360H}. 

\subsection{System Multiplicity}\label{sec:mult}
\subsection{Host Stars}
Twenty-one of the new co-moving systems reported here-in have primary stars with hints of binarity due to large Gaia RUWE values.

In addition to just the single RUWE from Gaia DR3, it is also helpful to compare the RUWE value from Gaia DR2, as DR3 has a longer baseline of observations \citep{2021A&A...649A...2L}. If a binary system has a long orbital period, there may not be enough time in the relatively short Gaia observation window to detect any deviation away from the single star model. However, if the orbital period was just right, the extra observations between DR2 and DR3 may be enough for the non-single star behavior to show up, causing an increase in the RUWE. Similarly, the extra observation time could result in the photocenter wobble averaging out, resulting in a decrease of the RUWE. 

\begin{figure*}
    \centering\includegraphics[width=1\textwidth]{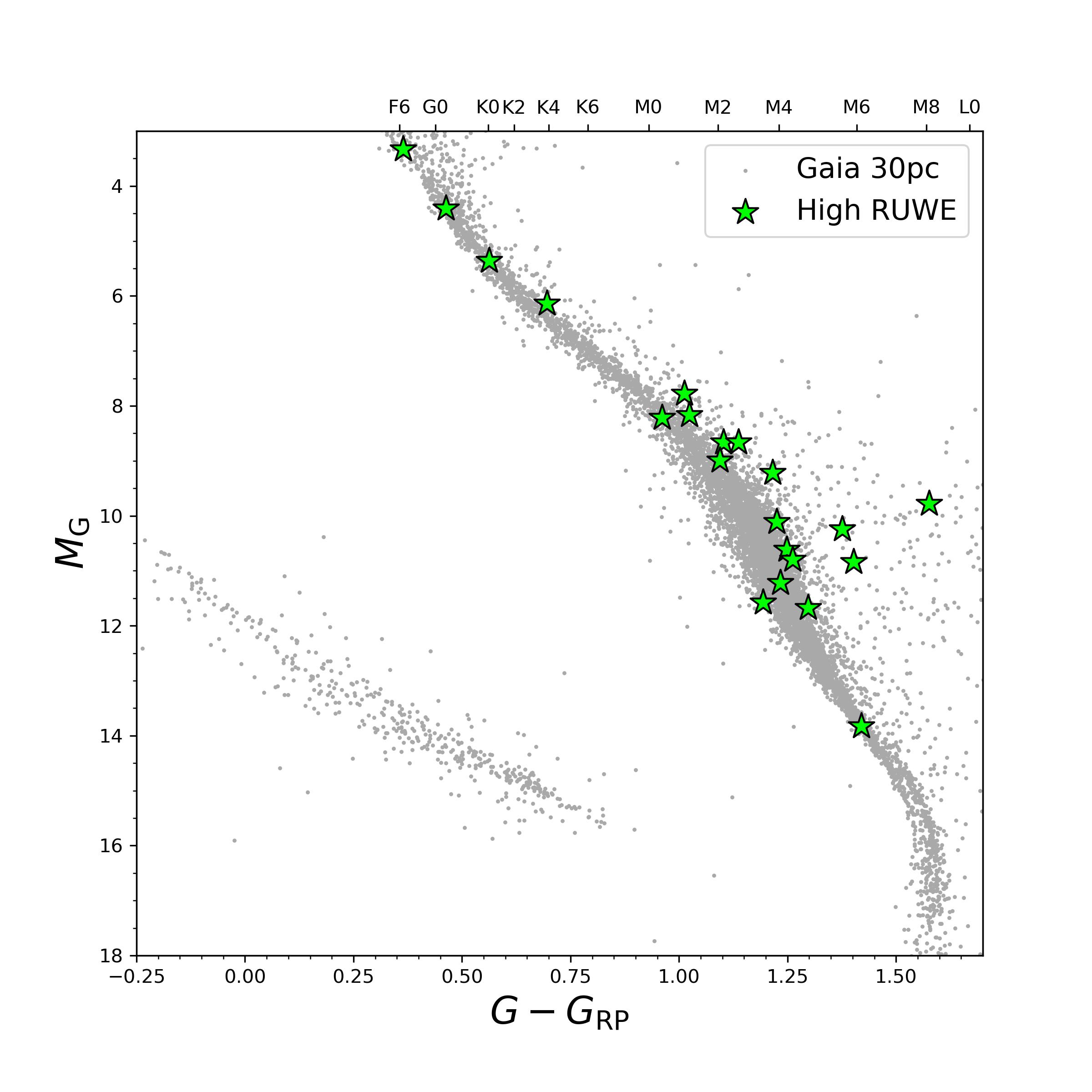}
    \caption{Host stars with Gaia DR3 RUWE values $>$1.4 as green stars, plotted against the Gaia DR3 30 pc sample.}
    \label{fig:8}
\end{figure*}

We therefore compared the RUWE values from DR2 and DR3 of all the primary stars of our co-moving systems. Using the nominal RUWE cutoff of $\sim$1.4, we find that twenty primaries were above this limit in Gaia DR3. These sources, as well as the hosts which are confirmed binaries, are listed in Table \ref{table: 8}. Looking at the locations of these stars on the Gaia CMD shown in Figure \ref{fig:8}, we find that a large majority of them appear to be overluminous, consistent with the interpretation of these sources being binaries. A few of the stars with a high RUWE value appear to be near normal on the Gaia CMD for their spectral type. While this could be indicating other sources for their anomalous RUWE values than binarity, this could also be due to a potential hidden companion faint enough to not contribute much to the total flux of the host, yet bright enough to effect the objects photocenter.

\subsection{Companion Multiplicity} \label{mult}
\subsubsection{Overluminous Candidates}
In addition to considering the binarity of the host stars, we can also take a look at the multiplicity of the companions. Ultracool dwarfs in wide binary pairs have been shown to have higher occurrence rates of binarity \citep{2005AJ....129.2849B,Faherty10, 2010ApJ...720.1727L}. Looking at our companions on the WISE CMD in Figure \ref{fig:1}, adopting the parallax of their hosts, we find that seven appear to be overluminous for their spectral type. Six of the companions that appear overluminous are estimated to have spectral types T5 or later, and have projected separations from their hosts $>1000$ AU. The wide separations and low mass ratios lead to very small binding energies for these systems (see Section \ref{sec:be}). Assuming the overluminosity of the companions is due to binarity, the binding energies of these systems could increase and begin to more closely resemble the binding energies observed in typical binary systems. 

\subsubsection{Spectral Binary Candidates}\label{sec:sb_diss}
Additionally, as shown in Table \ref{tab:sb}, 9 of the potential companions in the BYW benchmark sample are candidate spectral binaries.

\textbf{Strong Candidates} Two of our candidate companions are strong spectral binary candidates: CW0620+3446 (best single fit L4, best SB fit L4+T7) and CW2029-7910  (best single fit L1 blue, best SB fit sdL1+L2.5). As shown in Table \ref{table: 8}, CW2029-7910's host star, SCR J2029-7910, is itself a candidate binary. If the binary nature of both components are confirmed, this would make the CW2029-7910 + SCR J2029-7910 a widely separated quadruple system. 

\textbf{Weak Candidates} Seven of our candidate companions are weak spectral binary candidates: CW0227+0830 (best single fit L5 blue, best SB fit L5+L6), CW0627-0028 (best single fit T0 blue, best SB fit L5+T2.5), CW1313+2230 (best single fit L3, best SB fit L2+T1.5), CW1328-0637 (best single fit L3 blue, best SB fit L2+L5), CW1439+2600 (best single fit T3, best SB fit T1.5+T4.5), CW1606-1032 (best single fit L7 blue, best SB fit L4.5+L6.5), and CW2141-0246 (best single fit L4 sl. red, best SB fit L2+L3). \cite{2014ApJ...794..143B} note that blue L-dwarfs are a large contaminant in the analysis of weak spectral binary candidates, so blue sources such as CW1328-0637 and CW1606-1032 are potentially contaminants. The other weak candidates require further study in order to confirm their binarity.

\subsection{Binding Energy}\label{sec:be}
The binding energies of our systems were calculated with the equation: 

\begin{equation}
    1.8 \times 10^{5} * (\frac{(M_{companion} \times M_{host})}{Separation}
\end{equation}

where $M_{companion}$ and $M_{host}$ are given in units of $M_{\sun}$, the separation given in AU, and the constant chosen to give binding energy in units of $10^{41}$ ergs. The masses of our host stars were found as described in Section \ref{sec:mass_gather}. For the companions, when no age was available, we assumed them to be field age ($\sim$2-10 Gyr) and  used the spectral type dynamical mass bins of \cite{2017ApJS..231...15D}. For several of our systems, the host allowed for an age range to be estimated. For these systems, we used the evolutionary tracks of \cite{2008ApJ...689.1327S} to give a range of possible masses. In order to account for different orbital inclinations and eccentricities, we multiplied the projected separation by 1.26 (see \citealt{1992ApJ...396..178F}). Additionally, we estimated the dissipation lifetime of each system using the approach of \cite{2010AJ....139.2566D}, which estimates the length of time a system at a given separation would survive, based on average gravitational interactions within the galaxy. A summary of the component masses, mass ratios, binding energies, and estimated lifetimes for all systems is shown in Table \ref{table: 9}. 

\begin{figure*}
    \centering
    \includegraphics[width=1\textwidth]{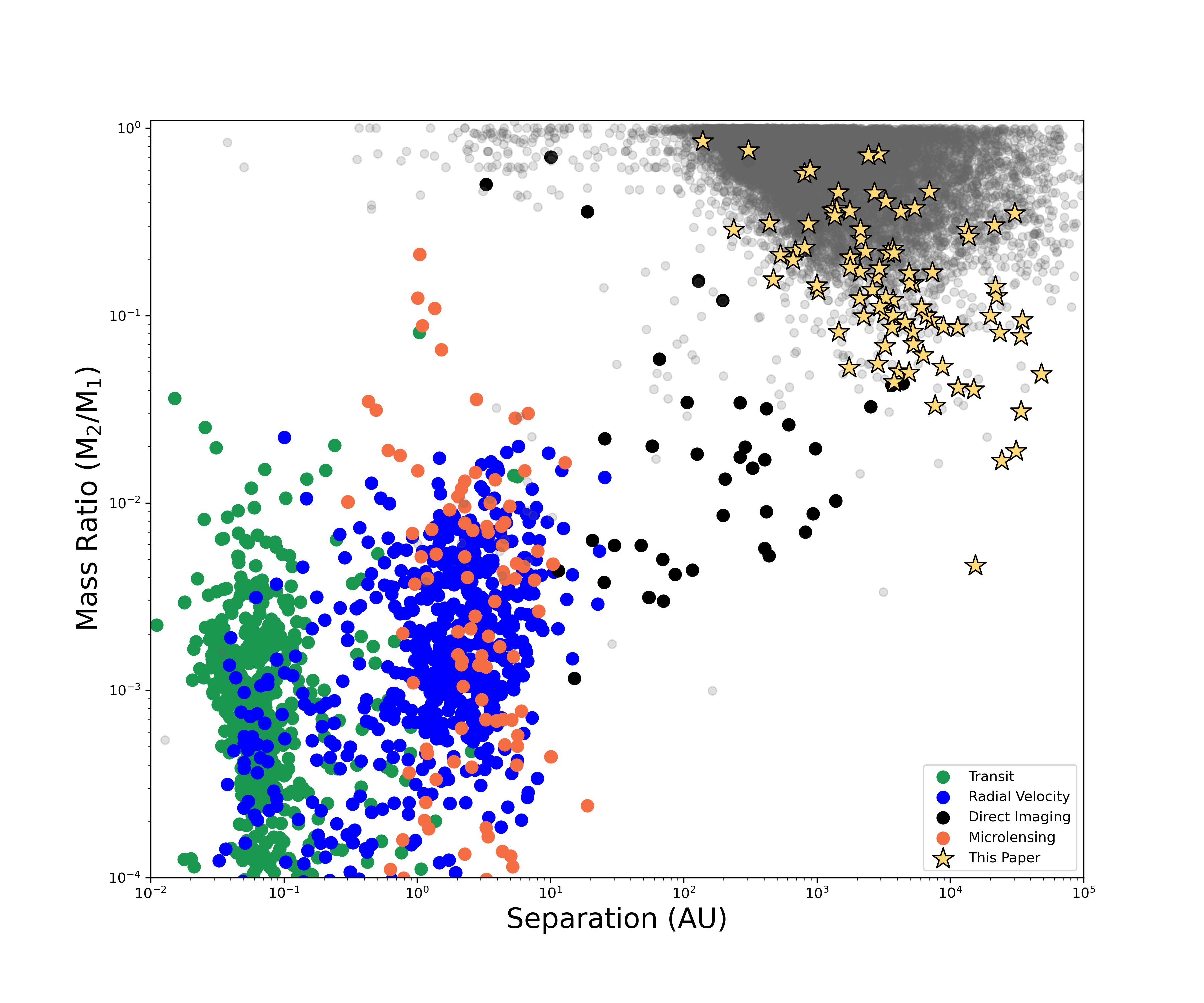}
    \caption{Separation vs. mass ratio for all systems in this sample as yellow pointed stars. Shown for comparison are Gaia stellar binaries, as well as exoplanets color coded by discovery method as listed on the Exoplanet Archive.}
    \label{fig:10}
\end{figure*}

Comparing the physical parameters of our benchmark systems with those of stellar binary and exoplanet systems may provide clues to their possible formation mechanisms. Figure \ref{fig:10} shows the separation vs.\ mass ratio for this sample compared against known stellar-stellar, brown dwarf-brown dwarf, and stellar-exoplanet systems. The exoplanet and stellar binary systems populate two distinct regions in Figure \ref{fig:10}, with the directly imaged exoplanets forming a bridge between the two populations. The majority of our sample lies amongst the stellar binary population in Figure \ref{fig:10}, with many filling in the region which connects with the widest and highest mass ratio directly imaged exoplanets. Several of the systems in our sample, despite having mass ratios more resembling directly imaged exoplanets, are found at extremely large separations. These systems, which occupy a region of Figure \ref{fig:10} which is nearly unpopulated, may be indicating the possibility for companion multiplicity, which agrees with previous studies as discussed in Section \ref{mult}.  

\begin{figure*}
    \centering
    \includegraphics[width=1\textwidth]{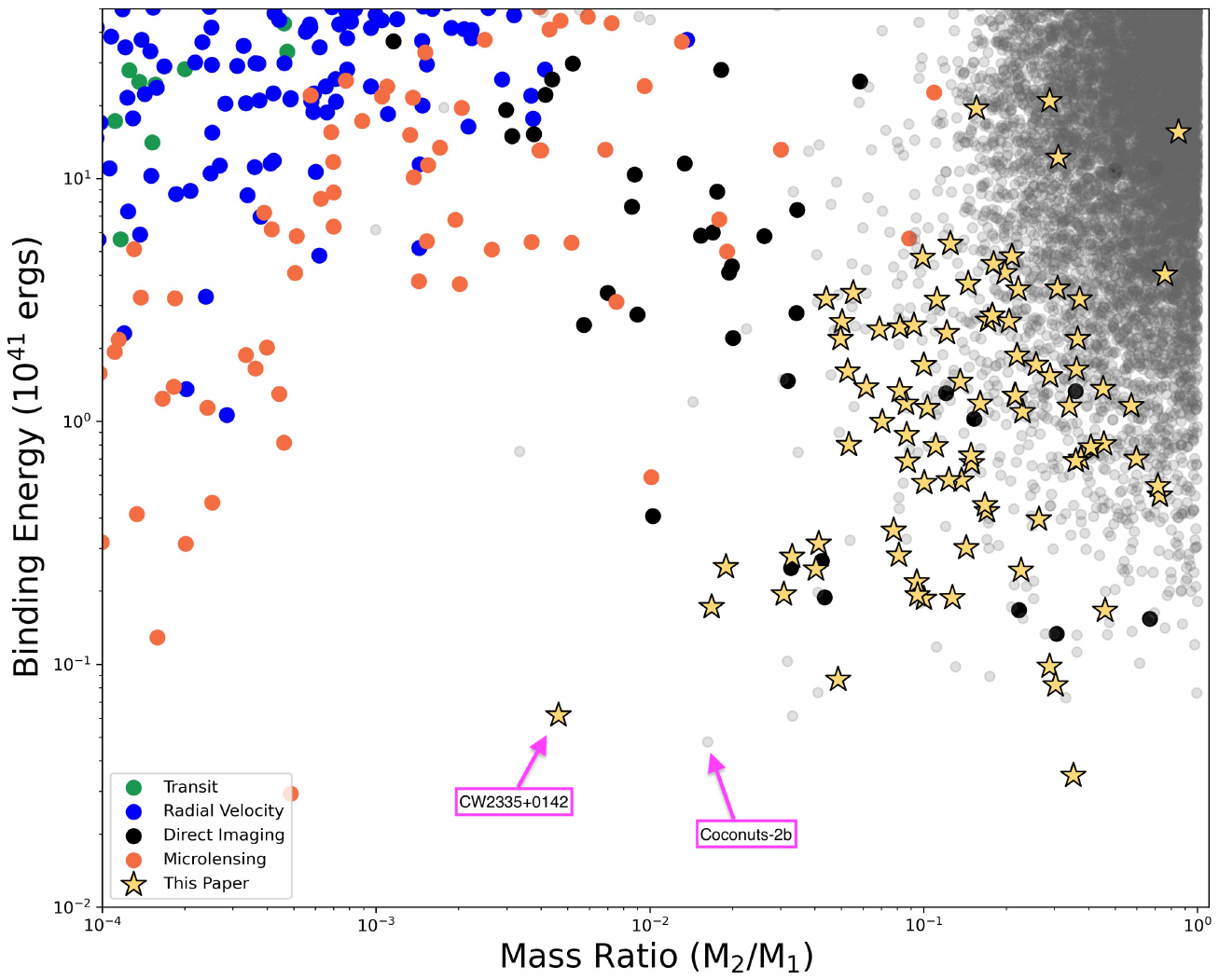}
    \caption{Mass ratio vs. binding energy for all systems in this sample as yellow pointed stars. Shown for comparison are Gaia stellar binaries, as well as exoplanets color coded by discovery method as listed on the Exoplanet Archive. CW2335+0142 is labeled, along with COCONUTS-2b for reference.}
    \label{fig:11}
\end{figure*}

Figure \ref{fig:11} shows the mass ratio vs.\ binding energy for the same sample. We can see that the systems in this sample are beginning to fill in the sparsely populated regions of low mass ratio and low binding energy which connect the directly imaged exoplanets to the stellar binary population. Several systems on Figure \ref{fig:11} begin to separate themselves from the other populations, with binding energies much lower than the majority of exoplanet or stellar binary systems.  

We find that five of our hypothesized systems have binding energies below 1 x $10^{40}$ ergs, the estimated minimum binding energy of substellar binary systems \citep{2003ApJ...586..512B,2007ApJ...660.1492C}, with another twenty systems approaching this limit, with binding energies $<$5 x $10^{40}$ ergs. These lowest binding energies may be suggestive for the need of higher multiplicity in these systems. For example, the lowest binding energy system of these is SDSS J134403.83+083950.9 + NLTT 35024, a T0+M5.5 pair with a separation of $\sim$24,100 AU and a binding energy of 3.49 x $10^{39}$ ergs. NLTT 35024 however is listed as an astrometric binary in Gaia DR3's non-single star solutions. Additionally, SDSS J134403.83+083950.9 was identified as a low probability L5+T2 binary in \cite{2015MNRAS.449.3651M}, raising the possible mass of the system even more. For more discussion on the possibility of higher multiplicity in this sample, please refer to Section \ref{sec:mult}.  

These low binding energies may also allow an age constraint to be placed on the system. If the system is young, there is a possibility that gravitational interactions have not yet had time to dissipate the pair. Indeed, several of the lowest binding energy substellar systems known are found at young ages, with very few found at field age (see \citealt{2007ApJ...660.1492C}), seeming to agree with the picture of wide, low binding energy systems dissipating over time through interactions with their natal environments. Previous studies have attempted to look at ultracool companion separations as a function of age (see \citealt{2011AJ....141...71F}) to investigate if there is indeed a correlation with wide separation and youth, but found no strong correlation. The wide, low binding energy systems in this sample are a new addition to the picture and provide an exciting population for future analysis on the widest ultracool companions. 

Further study of these low binding energy systems is required to identify any signs of higher multiplicity in either the host or companion, as well as for any youth diagnostics for a given system.

\section{Conclusions} \label{sec:conclude}
We have presented in this paper 89 new ultracool companions, including companions to main sequence stars identified via the citizen science project Backyard Worlds: Planet 9. We have obtained near-infrared spectra for 32 of the companions, along with an additional 3 near-infrared spectra of ultracool dwarfs which were originally identified as co-moving with a star, but have since been downgraded to a low likelihood of companionship due to their low scores with \texttt{CoMover}. We have also obtained both near-infrared or optical spectra for 33 of the hosts. The spectroscopic sample of companions shows a variety of different characteristics, including 7 young, 6 abnormally blue, and 3 peculiarly red brown dwarfs.

Comparing our BYW benchmark sample with known companions in the literature, we find that our sample adds 21 systems with an FGK primary, increasing the population of ultracool companions to FGK stars by $\sim$42\%. We also add 72 systems with projected separations $>$1000 AU, increasing the known population of ultracool companions with such wide separations by 154\%. The formation of these systems remains difficult to explain, although could point towards outward migration due to a third companion or possible dissolving moving groups. Further study of these systems should shed light on the formation and evolution processes of giant exoplanets and low mass stars, as they are beginning to fill in sparsely populated spaces on separation-binding energy diagrams.

Utilizing the properties of their host star to benchmark their properties, the ultracool dwarfs in this BYW benchmark sample provide a wealth of information, which can be utilized to test and constrain the atmospheric and evolutionary models of brown dwarfs and exoplanets. Seventy-two of the ultracool dwarf companions which are below the magnitude limit for astrometry with Gaia are able to adopt the milliarcsecond precision parallax of their host, allowing them to serve as parallax benchmarks needed for calibration of absolute magnitude - spectral type trends. Thirty-seven hosts in the BYW benchmark sample provide an age estimate for the system, breaking the mass-temperature-age degeneracy which plagues their ultracool dwarf companion. Twenty-four hosts in the BYW benchmark sample are FGK stars, which allow for a measure of their elemental abundances and composition. 

The BYW benchmark sample is a valuable resource for future retrieval studies, where it is crucial to be able to compare the retrieved abundances of the companion with the abundances of the host star. For example, this sample provides excellent targets for comparing the C/O ratios between host and companion, better understanding and constraining the silicate cloud properties of ultracool atmospheres using their host's Mg/Si ratios, or further constraining the role metallicity plays in changing the SEDs and atmospheric properties of ultracool dwarfs.

\begin{acknowledgments}
The authors would like to thank the numerous citizen scientist volunteers of the Backyard Worlds: Planet 9 project for their hard work, dedication, and contributions to this paper. 

The authors acknowledge support from NASA award \#80NSSC21K0402 and NASA ADAP award \#80NSSC22K0491 as well as NSF award \#2009177.  JF acknowledges support from the Heising Simons foundation.
This paper includes data gathered with the 6.5 meter Magellan Telescopes located at Las Campanas Observatory, Chile. Based in part on observations obtained at the Southern Astrophysical Research (SOAR) telescope, which is a joint project of the Minist\'{e}rio da Ci\^{e}ncia, Tecnologia e Inova\c{c}\~{o}es (MCTI/LNA) do Brasil, the US National Science Foundation’s NOIRLab, the University of North Carolina at Chapel Hill (UNC), and Michigan State University (MSU). 
Some of the data presented herein were obtained at Keck Observatory, which is a private 501(c)3 non-profit organization operated as a scientific partnership among the California Institute of Technology, the University of California, and the National Aeronautics and Space Administration. The Observatory was made possible by the generous financial support of the W. M. Keck Foundation. The authors wish to recognize and acknowledge the very significant cultural role and reverence that the summit of Maunakea has always had within the Native Hawaiian community. We are most fortunate to have the opportunity to conduct observations from this mountain. Some of the observations reported in this paper were obtained with the Southern African Large Telescope (SALT). This work presents results from the European Space Agency (ESA) space mission Gaia. Gaia data are being processed by the Gaia Data Processing and Analysis Consortium (DPAC). Funding for the DPAC is provided by national institutions, in particular the institutions participating in the Gaia MultiLateral Agreement (MLA). The Gaia mission website is https://www.cosmos.esa.int/gaia. The Gaia archive website is https://archives.esac.esa.int/gaia. This publication makes use of data products from the Wide-field Infrared Survey Explorer, which is a joint project of the University of California, Los Angeles, and the Jet Propulsion Laboratory/California Institute of Technology, funded by the National Aeronautics and Space Administration.
\end{acknowledgments}

\appendix

Several systems in our BYW benchmark sample show spectral or kinematic indications of youth, low metallicity, or anomalous colors. Below we enumerate on the properties of these interesting sources and their host stars.

\section{Young Systems}
 \label{sec:noteworthy}
\subsection{UCAC3 113-933 + CWISE J002101.45-334631.5}
The candidate host star UCAC3 113-933 is estimated to have a spectral type of  M0 (using the Gaia photometry and absolute magnitude relations of \citealt{2019AJ....157..231K}). UCAC3 113-933 was observed in the RAVE survey, and has an age estimate in the literature of $\sim$155 Myr \citep{2017ApJ...835...61Z}. 
Given the caveats of this study, we adopt an age of 0.1--1 Gyr for UCAC3 113-933. Figure \ref{fig:1} shows that UCAC3 113-933 does not appear overluminous on the Gaia CMD compared to stars of the same color. UCAC3 113-933 has a corresponding UV source with a near UV magnitude of NUV=21.08$\pm$0.18 mag \citep{2017ApJS..230...24B}. No X-ray, H$\alpha$, or rotational information is available in the literature. Running UCAC3 113-933's full kinematics (proper motions, radial velocity, and parallax) through BANYAN $\Sigma$, we find a field probability of 99.9$\%$. UCAC3 113-933 has an anomalously large Gaia DR3 RUWE of 1.98, hinting at the possibility of a hidden companion.

CWISE J002101.45-334631.5 (CW0021-3346) is a common-proper motion companion to UCAC3 113-933.
No spectrum of CW0021-3346 exists. The infrared colors of CW0021-3346 (W1-W2=2.35$\pm$0.16 mag, $J$-$H$=0.31$\pm$0.25 mag, $J$-W2=4.38$\pm$0.15 mag) indicate a mid to late T dwarf; while
the W2 absolute magnitude (derived via adopting the parallax of UCAC3 113-933) yields a spectral type estimate of (T1) and places the source as slightly more luminous than typical brown dwarfs of the same color on the WISE CMD (Figure \ref{fig:1}). While it has been shown that young L dwarfs appear fainter and redder than their field counterparts \citep{2012ApJ...752...56F,2016ApJ...833...96L}, the mid-infrared absolute magnitudes of young T dwarfs are more comparable to the older, field population (\citealt{2020ApJ...891..171Z}).  This raises the possibility that the excess brightness of CW0021-3346 is due to binarity. Visually inspecting the $Z$-, $Y$-, $J$-, $H$-, and $K$-band images available from the VISTA Kilo-degree Infrared Galaxy (VIKING) survey \citep{2013Msngr.154...32E}, we find no evidence of a companion, although its faint near-infrared magnitudes (J=19.9 mag) make a clear determination difficult. Additionally, the potential pair may be too close to be resolved, leaving the possibility of binarity still open. As has been noted in \cite{Faherty10}, widely separated ultracool dwarfs have a higher likelihood of being a close double themselves.  Therefore binarity is a viable option for explaining the position of CW0021-3346 in Figure \ref{fig:1}.

The angular separation of this system is $\sim$116\arcsec. At the distance of UCAC3 113-933 the pair have a projected physical separation of $\sim$6100 AU. With both components in this system appearing brighter than their field counterparts, taking into consideration UCAC3 113-933's young age range from \cite{2017ApJ...835...61Z}, we believe the UCAC3 113-933 + CW0021-3346 system to be a 0.1--1 Gyr (M0)+(T5) widely separated binary.   

\subsection{CWISE J022454.80+152629.5 + CWISE J022454.10+152633.8}
CWISE J022454.80+152629.5 (CW0224+1526A) is classified as an L6 (red) based on the comparison of its near infrared spectrum with that of the L6 spectral standard, shown in Figure \ref{fig: 3}. CW0224+1526A's spectrum appears slightly redder than the L6 spectral standard, and is indicative of a slightly more triangular H-band, suggestive of youth. We calculated the gravity sensitive indices of \cite{2013ApJ...772...79A} for CW0224+1526A (Table \ref{table: 4}) and find assign it a gravity classification of INT-G. Running the kinematics of CW0224+1526A through BANYAN $\Sigma$, and find that CW0224+1526A receives a 89.9$\%$ probability of membership to the $\beta$ Pictoris moving group, a 10$\%$ field probability, and a 0.1$\%$ probability of AB Doradus moving group membership. 

CWISE J022454.10+152633.8 (CW0224+1526B) is a common proper motion companion with CW0224+1526A. The infrared colors of CW0224+1526B (J-W2=4.41$\pm$0.22 mag, W1-W2=0.80$\pm$0.06 mag) are indicative of an object near the L/T boundary, with the W1-W2 color implying a spectral type estimate of (T1). We ran the kinematics of CW0224+1526B through BANYAN $\Sigma$ and find a 72.6$\%$ probability of membership to the $\beta$ Pictoris moving group, a field probability of 23.1$\%$, a 2.4$\%$ probability of AB Doradus membership, and a 1.2$\%$ probability of Tucana-Horologium membership.

CW0224+1526A and CW0224+1526B are separated on the sky by an angular separation of 11\arcsec. Using the absolute W2 - spectral type relations of \cite{2021ApJS..253....7K}, we estimate CW0224+1526A to be at a distance of $\sim$36 pc, which translates to a projected physical separation for CW0224+1526AB of 400 AU. Both CW0224+1526 have higher probabilities of membership to the $\beta$ Pictoris moving group, which has an age of 23$\pm$3 Myr \citep{2014MNRAS.445.2169M}. However, it would be expected for objects at the age of $\beta$ Pictoris to have gravity classifications of VL-G, so the INT-G classification of CW0224+1526A may be hinting that the CW0224+1526AB pair may in fact be an interloper. With the pairs kinematics still consistant with known young moving groups, along with the red spectral morphology and INT-G classification of CW0224+1526A, we believe CW0224+1526AB to be a $<$1 Gyr widely separated L6 (red)+(T1) binary. 

\subsection{HD 51400 + CWISE J065752.45+163350.2}
HD 51400, is a solar-type star with a spectral type of G5 \citep{1993yCat.3135....0C}. On the Gaia CMD in Figure \ref{fig:1}, HD 51400 appears normal for its spectral type. Using the rotational period of HD 51400 from its TESS light curve, and placing it on the color-period diagram in Figure \ref{fig: 9}, it appears to have an age $\gtrsim$650 Myr. No information on X-ray, UV, or H$\alpha$ emission is listed in the literature. HD 51400 has an anomalously large Gaia DR3 RUWE of 22.44, and is listed as an astrometric binary in Gaia DR3's Non-Single Star Solution tables \citep{2023A&A...674A..34G}. This strongly suggests that HD 51400 is in fact a binary itself. Running the 3D kinematics of HD 51400 through BANYAN $\Sigma$, we find a field probability of 99.9$\%$.

CWISE J065752.45+163350.2 (CW0657+1633) is a common proper motion companion to HD 51400. We classify CW0657+1633 as spectral type L6 based on its near-infrared spectrum (Figure \ref{fig: 3}). Calculating the gravity sensitive indices of \cite{2013ApJ...772...79A}, we find that CW0657+1633 has a gravity classification of INT-G, indicating a surface gravity lower than field age objects of the same spectral type. On the WISE CMD (Figure \ref{fig:1}), using the parallax of its suspected primary, CW0657+1633 appears slightly fainter than typical objects of a similar spectral type. Running its proper motions through BANYAN $\Sigma$, as well as an estimated distance based on its spectral type, we find that CW0657+1633 has a field probability of 99.9$\%$.

HD 51400 and CW0657+1633 have an angular separation of 61\arcsec. At the Gaia distance of HD 51400, this translates to a projected physical separation of $\sim$2,300 AU. Due to the age estimate of HD 51400 from its rotation rate, as well as the INT-G gravity score of CW0657+1633, we believe HD 51400 + CW0657+1633 to be a 0.65 - 1 Gyr, widely separated G5+L6 system.  

\subsection{CWISE J101533.05-111501.0AB + CWISE J101523.92$-$111539.6}
CWISE J101533.05-111501.0A (1015-1115A) and CWISE J101533.05-111501.0B (1015-1115B) form a close ($\sim$2\arcsec) binary pair. We obtained an optical spectrum of 1015-1115A, and classify it as an M5 (Figure \ref{fig: 6}). Using the Gaia photometry and absolute magnitude relations of \cite{2019AJ....157..231K}, we estimate 1015-1115B to have a spectral type of M5 as well. On the Gaia CMD (Figure \ref{fig:1}), we find that while 1015-1115B appears normal for its color, 1015-1115A appears to be overluminous. In addition, the spectrum of 1015-1115A in Figure \ref{fig: 6} shows noticeable H$\alpha$ emission. There is a near UV source associated with the binary pair (NUV=22.51$\pm$0.42 mag; \citealt{2017ApJS..230...24B}), although there is no associated X-ray emission. We find a rotational period of P$_{rot}$=0.35$\pm$0.11 days from the TESS light curve, however due to the large pixel size of TESS both 1015-115A and 1015-1115B appear in the same pixel. Making the interpretation of this rotation period more complicated is that 1015-1115A has a large Gaia DR3 RUWE value of 1.67, indicating the possibility of a hidden companion. On the color-period diagram shown in Figure \ref{fig: 9}, 1015-1115A appears to be consistent with an age of only a few hundred of millions of years. We ran 1015-1115A through BANYAN $\Sigma$ (as 1015-1115A was the only star in the pair with full kinematics in Gaia) and find that it receives a probability of membership to the TW Hydrae Association (TWA) of 85.2$\%$ and a field probability of 14.8$\%$ .

CWISE J101523.92$-$111539.6 (CW1015-1115) is a common proper motion companion with the 1015-1115A and 1015-1115B pair. We assign CW1015$-$1115 a spectral type of L5 based on its near infrared spectrum in Figure \ref{fig: 3}. It can be seen that the spectrum of CW1015$-$1115 appears slightly redder compared to the L5 standard in Figure \ref{fig: 3}, with an excess of flux in the $H$- and $K$-bands. The peak of the $H$-band of CW1015$-$1115's spectrum also displays a distinctly more triangular shape. Using the gravity sensitive indices of \cite{2013ApJ...772...79A} (listed in Table \ref{table: 4}), we find that CW1015$-$1115 receives a gravity score of INT-G, indicating a lower surface gravity than the field population. CW1015$-$1115's location on the WISE CMD in Figure \ref{fig:1} is in good agreement with other mid L dwarfs, although it appears slightly brighter than the field L5s. Young L dwarfs appear fainter in the near-infrared than their field counterparts, though have been shown to be overluminous in mid-infrared bands \citep{2016ApJ...833...96L}, so CW1015-1115's elevated position also hints at a possible younger than field age. This overluminosity may instead be indicative of CW1015-1115 being a close binary itself, however, which is thought to be more common in widely separated systems such as this \citep{Faherty10}. Running the kinematics of CW1015$-$1115 through BANYAN $\Sigma$, using an estimated parallax of 29$\pm$6 mas (derived from spectral type - M$_{W2}$ relations), we find a membership probability of 86.1$\%$ for TWA and a field probability of 13.9$\%$, similar to the results for 1015-1115A. Running BANYAN $\Sigma$ on CW1015-1115 without an estimated parallax results in a 71.2$\%$ probability of TWA membership, a 1.5$\%$ probability of membership to the Carina Near Moving Group, and a 27.3$\%$ field probability.  

CW1015$-$1115 has an angular separation from 1015-1115A and 1015-1115B of $\sim$140\arcsec. At the 1015-1115A, this translates to a projected physical separation of $\sim$5,800 AU. While the system receives a high membership probability of TWA membership, the INT-G gravity score of CW1015$-$1115 is at odds with this assessment. \cite{2016ApJS..225...10F} found that all TWA members of spectral type M7 or later received gravity scores of VL-G. While not impossible, as other physical processes may be impacting the relevant spectral features, it is unlikely for CW1015-1115 to receive an INT-G gravity score at the age of TWA, and may be an interloper to the YMG. Future follow up is required to further assess the spectral and kinematic indications of youth for CW1015-1115. However, due to 1015-1115A's H$\alpha$ emission and fast rotation rate, the redder color,  triangular H band, and INT-G gravity score of CW1015$-$1115's spectrum, we believe the 1015-1115A + 1015-1115B + CW1015$-$1115 to be a 0.1-1 Gyr, widely separated M5+(M5)+L5$\beta$ triple. 

\subsection{LSPM J1417+0418 + CWISE J141737.21+041847.2}
LSPM J1417+0418 is a known high proper motion star. While both the Gaia photometry - spectral type relations of \cite{2019AJ....157..231K}, as well as the Gaia CMD location in Figure \ref{fig:1}, place LSPM J1417+0418 as a mid-M dwarf, visual inspection using the WiseView tool leads us to believe this to be a result of contamination. WiseView includes a feature which allows for an overlay to be placed on the image, representing the parallax and proper motion of Gaia sources within the field of view. In looking at the location of LSPM J1417+0418 using this feature, there does not appear to be a moving WISE source associated with it, only a stationary source. This lack of an associated infrared source to a Gaia source is sometimes seen with white dwarfs, which are not particularly bright in these wavelengths.
Inspecting higher resolution images from DSS, Pan-STARRS, and UKIDSS, there do indeed appear to be two different sources: one which is believed to be a background galaxy, SDSS J141737.75+041823.9 \citep{2011AJ....141..189V}, and one with a similar proper motion as the Gaia source. While the extended source nature of the background object potentially contaminates the photometry of LSPM J1417+0418, making any spectral type estimating difficult, we believe LSPM J1417+0418 to be a possible white dwarf based on the lack of an associated moving infrared source. Further observations of LSPM J1417+0418 are required in order to confirm. As the Gaia RUWE for LSPM J1417+0418 of RUWE=1.29 falls below the nominal 1.4 limit for a 'well behaved' single star (see \citealt{2021ApJ...907L..33S}), the Gaia astrometry for LSPM J1417+0418 is likely unaffected by the contamination. Running the proper motions and parallax for LSPM J1417+0418 through BANYAN $\Sigma$, we find a 99.9\% field probability. 

CWISE J141737.21+041847.2 (CW1417+0418) is a common proper motion companion with LSPM J1417+0418. CW1417+0418 is found to also have a corresponding Gaia source, whose parallax of $\varpi$=16.01$\pm$0.45 mas is similar to the parallax of LSPM J1417+0418 of $\varpi$=17.07$\pm$0.27. The near infrared spectrum of CW1417+0418 is best matched by the near infrared spectrum of the M8 standard (Figure \ref{fig: 3}). Using the indices of \cite{2013ApJ...772...79A} we find that CW1417+0418 receives a gravity score of INT-G (listed in Table \ref{table: 4}), indicating an object with a surface gravity lower than the field population. Looking at CW1417+0418 on the WISE CMD in Figure \ref{fig:1}, we see that it appears near normal for its spectral type. On the Gaia CMD in Figure \ref{fig:1} however, CW1417+0418 appears slightly redder than objects of the same spectral type, placing it just to the right of the late M main sequence. Running the Gaia proper motions and parallax of CW1417+0418 through BANYAN $\Sigma$, we find a 99.9\% field probability. 

LSPM J1417+0418 and CW1417+0418 are separated on the sky by an angular separation of $\sim$24\arcsec. At the Gaia distance of LSPM J1417+0418, this translates to a projected physical separation of $\sim$1,400 AU. While the gravity sensitive indices for CW1417+0418 are suggestive of an object with intermediate gravity, the unknown nature and age of its host star makes an accurate determination difficult. It is unlikely that LSPM J1417+0418 is young, however, if its white dwarf nature is confirmed. This hints that CW1417+0418's INT-G classification is not due to youth, but to other processes subtly affecting the observed spectrum as has been found for several other M dwarfs (c.f. TRAPPIST-1 \citealt{2019ApJ...886..131G}) We therefore believe LSPM J1417+0418 + CW1417+0418 to be a widely separated (WD)+M8$\beta$. 

\subsection{UCAC4 840-013771 + CWISE J162511.27+774946.8}
\begin{figure*}
    \centering
    \includegraphics[width=1\textwidth]{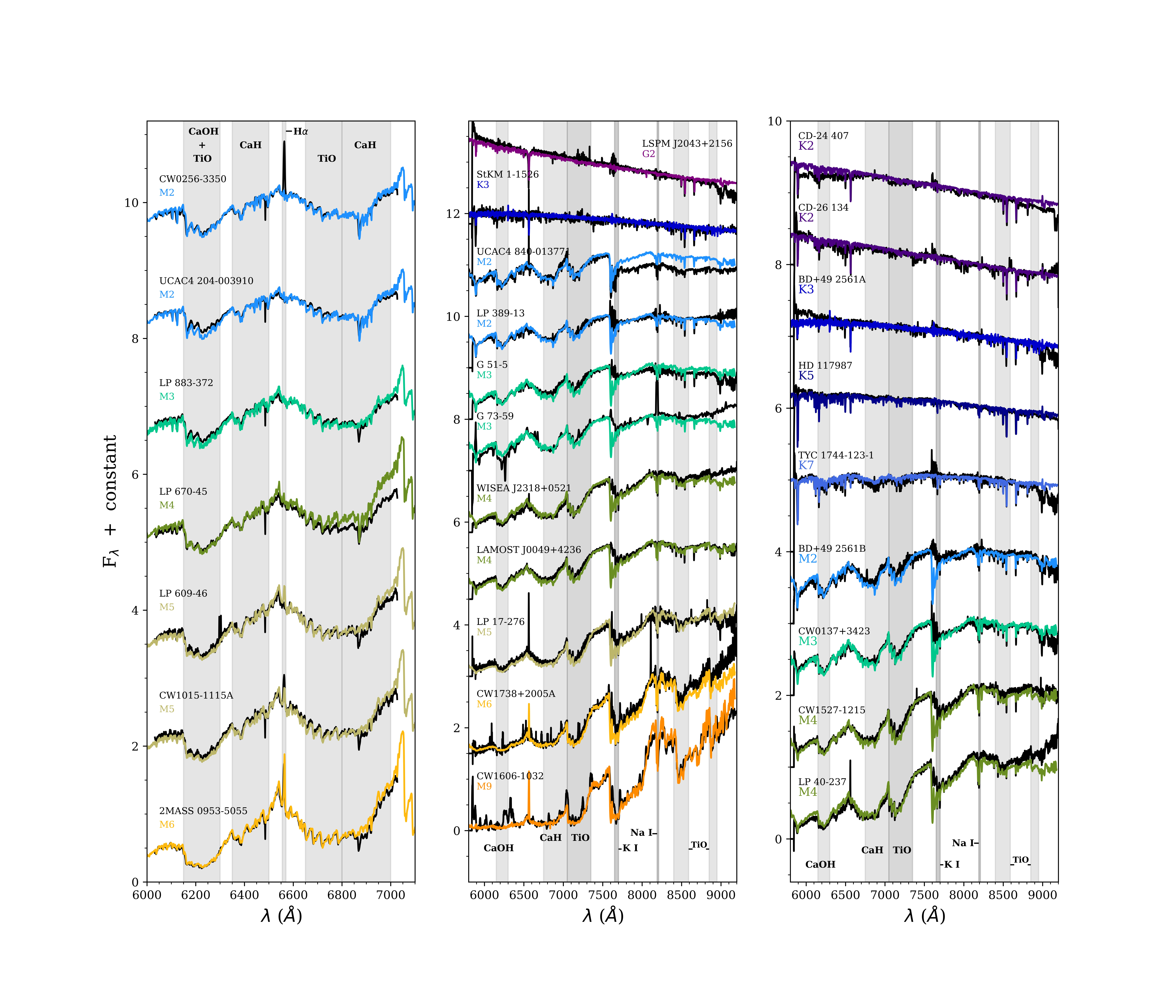}
    \caption{Optical spectra of the primary components plotted in black. Over-plotted in color are the spectral standards. All spectra were normalized between 7499-7501 \AA~ and separated by a constant.\\
    All spectra on left plot were obtained with SALT. Spectra in the middle and right plots were obtained with Lick/Kast.}
    \label{fig: 6}
\end{figure*}

UCAC4 840-013771 is a high proper motion star which we classify as spectral type M2 based on its optical spectrum, obtained from the CFHT archive (Figure \ref{fig: 6}). While UCAC4 840-013771's spectrum matches the M2 spectral standard between $\sim$6000 \AA\ and $\sim$7600 \AA\ in Figure \ref{fig: 6}, it is heavily suppressed compared to the spectral standard $>$7600 \AA. We note this suppression may be due to reduction or observation problems. Figure \ref{fig: 6} also shows the spectrum of UCAC4 840-013771 exhibits strong H$\alpha$ emission. \cite{2006AJ....132..866R} report an H$\alpha$ equivalent width of 4.5 \AA, agreeing with the strong H$\alpha$ emission in the observed spectrum. UCAC4 840-013771 is an X-ray and UV emitter, with an observed X-ray flux of F$_X$= 6.72 $\times$ 10$^{-13}$ erg s$^{-1}$ cm$^{-1}$ \citep{2022AA...664A.105F} and UV magnitudes of FUV=21.04$\pm$0.24 mag and NUV=19.99$\pm$0.09 mag \citep{2017ApJS..230...24B}. Using the TESS lightcurve of UCAC4 840-013771, we find a rotation period of P$_{rot}$=1.3 days. On the color-period diagram shown in Figure \ref{fig: 9}, UCAC4 840-013771 appears to be consistent with an age of at most a few hundred million years. The light curve of UCAC4 840-013771 also shows evidence of substantial flare activity. Looking at its position on the Gaia CMD in Figure \ref{fig:1}, UCAC4 840-013771 appears overluminous compared to other stars of the same color. The Gaia DR3 RUWE for UCAC4 840-013771 is larger than typical for single stars (RUWE=1.61), which could possibly indicate a hidden companion, or be a result of variability due to the bright flares evident in its TESS light curve. Running UCAC4 840-013771 through BANYAN $\Sigma$ with the full kinematics from Gaia DR3, we get a 99.9$\%$ field probability.

 CWISE J162511.27+774946.8 (CW1625+7749) is a common proper motion companion with UCAC4 840-013771. We classify CW1625+7749 as spectral type L4 from its near infrared spectrum (Figure \ref{fig: 3}). The Keck/NIRES spectrum for this source was taken through variable and thick clouds, and its high declination resulted in a large airmass difference between it ($z$ = 1.9) and the nearest A0~V telluric calibrator ($z$ = 1.5). As such, both the relative calibration across orders (e.g., the anomalously bright K-band) and the correction of telluric absorption (in particular the 1.1--1.5~$\mu$m band) are inaccurate. In order to properly spectral type CW1625+7749, we followed the procedure used in the past on extremely red L dwarfs (see for example \citealt{2014MNRAS.439..372M,2018AJ....155...34C}): each of the 3 bands were separated and individually normalized, then compared with the spectral standards treated in the same manner. This band-by-band classification results in a classification of L2. Calculating the indices of \cite{2013ApJ...772...79A} (Table \ref{table: 4}), we find that CW1625+7749 receives a gravity score of FLD-G, indicating an object of field age surface gravity. We note though that this score may not be accurate, due to potential reduction issues caused by strong telluric absorption and flux calibration issues. Looking at the position of CW1625+7749 on the WISE CMD, it does not appear to be under or over luminous compared with field objects of the same spectral type. We ran CW1625+7749 through BANYAN $\Sigma$, using an estimated parallax of 15.20$\pm$3.04 mas derived from spectral type - M$_{W2}$ empirical relations, and find a 99.9$\%$ field probability. 

 CW1625+7749 and UCAC4 840-013771 are separated on the sky by an angular distance of $\sim$313\arcsec. At the Gaia DR3 distance of UCAC4 840-013771, this translates to a projected physical separation of $\sim$19,000 AU. Given UCAC4 840-013771's elevated CMD, H$\alpha$, X-ray and UV emission, as well as its fast rotation rate and flares, we believe UCAC4 840-013771 + CW1625+7749 to be a 0.1--1 Gyr M2+L2 wide binary.

\subsection{TYC 5213-545-1 + CWISE J214129.80-024623.6}
TYC 5213-545-1 is classified as a K0 dwarf \citep{2019ApJS..245...34X}. Using the TESS light curve of TYC 5213-545-1, we find a rotational period of P$_{rot}$=14.3 days, with hints of multiperiodic oscillations. No H$\alpha$, X-ray, or UV data is reported in the literature. \cite{2017ApJ...835...61Z} report an age for TYC 5213-545-1 of $\sim$550 Myr based on the Ca II infrared triplet equivalent width excess, and we adopt an age range of 0.1--1 Gyr. On the Gaia CMD, TYC 5213-545-1 does not appear overluminous compared to stars of similar color. Its Gaia DR3 RUWE is near one (RUWE=0.95). Running the full Gaia kinematics of TYC 5213-545-1 through BANYAN $\Sigma$, we find a 99.9$\%$ field probability.

CWISE J214129.80-024623.6 (CW2141-0246) is a common proper motion companion with TYC 5213-545-1. We assign CW2141-0246 a spectral type of L4 based on its near infrared spectrum. In Figure \ref{fig: 3} the spectrum of CW2141-0246 appears more red than the standard L4, with its $H$ and $K$ bands slightly enhanced. The peak of CW2141-0246's $H$-band does not show the typical triangular shape, but is instead much flatter than the spectral standard in Figure \ref{fig: 3}. CW2141-0246 receives a gravity classification of FLD-G using the indices of \cite{2013ApJ...772...79A}, indicating an object with field age surface gravity. The WISE CMD in Figure \ref{fig:1} shows that CW2141-0246 is in good agreement with typical field dwarfs of the same color. Table \ref{tab:sb} shows CW2141-0246 is consistent with being a weak spectral binary candidate.  BANYAN $\Sigma$, with an estimated parallax of 15.9$\pm$3.2 mas derived from empirical spectral type - M$_{W2}$ relations, we find a field probability of 99.9$\%$.

TYC 5213-545-1 and CW2141-0246 are separated on the sky by $\sim$114\arcsec. At the Gaia DR3 distance of TYC 5213-545-1, this translates to a projected physical separation of $\sim$9,000 AU. Given the red spectrum of CW2141-0246, along with the age range of 0.1--1 Gyr from \cite{2017ApJ...835...61Z}, we believe TYC 5213-545-1 + CW2141-0246 to be a 0.1--1 Gyr K0+L4 wide binary.

\subsection{GJ 900 + CWISE J233531.55+014219.6}
\label{sec:2335}
GJ 900 is a multiple system, where the components A, B, and C have spectral types K7, M4, and M6 respectively \citep{2007AstBu..62..117M}. GJ 900 is a hierarchical triple: GJ 900A and GJ 900BC orbiting a common center of mass with a period of $\sim$80 years, and GJ 900BC form a close binary pair with an orbital period of $\sim$20 yrs . \cite{2007AstBu..62..117M} also report the detection of two fainter objects in 2MASS images which are theorized to potentially be late M dwarfs comprising components D and E of the GJ900 system. No follow-up observations have confirmed these sources as part of the GJ 900 system, and are likely not associated. On the Gaia CMD in Figure \ref{fig:1}, GJ 900 appears to be overluminous, as is expected for multiple systems. \cite{2018ApJ...856...23G} assign GJ 900 a bona-fide membership of the Carina-Near moving group (200$\pm$50 Myr). We reran GJ 900 through BANYAN $\Sigma$ using the updated kinematics from Gaia DR3 and find a 99.7$\%$ membership probability to Carina-Near and a 0.3$\%$ field probability, which is in good agreement with \cite{2018ApJ...856...23G}. Using its light curve from TESS, we find a rotational period of 11.92 days, along with a smaller amplitude periodicity of $\sim$8 days. The light curve of GJ 900 also shows evidence of flares. On the color-period diagram in Figure \ref{fig: 9}, GJ 900's position is consistent with its given age of 200$\pm$50 Myr. The GJ 900 system is a source of X-ray emission, with an observed flux of 9.13 $\times$ $10^2$ $mW/m^2$ \citep{2018MNRAS.473.4937S}, as well as a source of UV emission, with observed magnitudes of FUV=19.94$\pm$0.14 mag and NUV=17.52$\pm$0.03 mag \citep{2017ApJS..230...24B}. GJ 900 has multiple reported metallicity measurements in the literature ranging from [Fe/H] of --0.6 to +0.3  \citep{1996A&AS..117..227A, 2006ApJ...638.1004A, 2016ApJ...822...97H, 2016A&A...591A.118S, 2019AJ....158...56H, gaia2023}. 

CWISE J233531.55+014219.6 (CW2335+0142) is a common proper motion companion with GJ 900. The near infrared spectrum of CW2335+0142 in Figure \ref{fig: 3}, while low signal-to-noise due to observational challenges, is consistent with a cold object of late-T spectral type, and is best fit as spectral type T9. On the WISE CMD in Figure \ref{fig:1}, CW2335+0142 does not appear to be over- or under-luminous when compared to objects of a similar color. Running the kinematics and estimated photometric distance CW2335+0142 through BANYAN $\Sigma$, we find a 73.2$\%$ probability of $\beta$ Pictoris moving group membership, a 12.7$\%$ probability of AB Doradus Moving Group membership, 6.4$\%$ probability of Argus Moving Group membership, and a 7.6$\%$ field probability. While these membership probabilities for CW2335+0142 are not in agreement with those of its suspected host (99.7$\%$ Carina-near), this may be due to the large proper motion uncertainties ($\sim$85 mas/yr) used for CW2335+0142 in BANYAN $\Sigma$. However, with only a 7.6$\%$ field probability, the kinematics of CW2335+0142 are still strongly suggestive of it belonging to a young population. Further observations of CW2335+0142, including higher accuracy proper motions, are required in order to confirm any potential memberships. 

GJ 900 and CW2335+0142 are separated on the sky by an angular separation of $\sim$587\arcsec. At the distance of GJ 900, this translates to a projected physical separation of $\sim$12,000 AU. Despite the disagreement of YMG memberships from BANYAN $\Sigma$, the GJ 900 + CW2335+0142 pair still receive a 99.9$\%$ probability of being a co-moving pair via the \texttt{CoMover} code. Given the high co-moving probability from \texttt{CoMover}, along with GJ 900's membership of the Carina-Near moving group, we believe GJ 900 + CW2335+0142 to be a 200$\pm$50 Myr widely-separated (K7+M4+M6)+T9 multiple system.

One of the only comparably cold, widely separated substellar companions known is COCONUTS-2b, a T9 with and age of 150-800 Myr and a mass of 6.3$^{+1.5}_{-1.9}$ \mjup \citep{2021ApJ...916L..11Z}. If confirmed to belong to the Carina-Near moving group, CW2335+0142 would also be a planetary mass object, adding to the small number of widely separated, cold exoplanets.

\section{Unusually Blue Spectra}
The brown dwarf population has several subpopulations that showcase how varying secondary parameters might alter observable features. One such subpopulation is subdwarfs -- low metallicity objects which present enhanced collision induced H$_2$ absorption resulting in a bluer than normal near-infrared spectrum \citep{1969ApJ...156..989L,1994ApJ...424..333S}. 
In addition to low metallicity, in L dwarfs, abnormally blue near-infrared colors have been shown to be a possible indication of unresolved binarity with a T dwarf component \citep{2007ApJ...659..655B, 2014ApJ...794..143B}.
There are also some ultracool dwarfs which present bluer than normal spectra, but do not show spectral indications of low metallicity or binarity \citep{2000AJ....120.1085G,2003AJ....126.2421C,2010ApJS..190..100K}. 
It has been hypothesized that large grained or thin condensate clouds may be responsible for these blue L-dwarfs \citep{2008ApJ...674..451B}. Below we discuss the systems which presented bluer than expected near-infrared spectra.

\subsection{G 73-59 + CWISE J022737.75+083008.8}
\label{sec:0227}
G 73-59 is a high proper motion star, which 
our Kast optical spectrum indicates a spectral type of M3. On the Gaia CMD in Figure \ref{fig:1}, G 73-59 appears to have the same absolute magnitude as stars with a similar color. The Gaia DR3 RUWE of G 73-59 is abnormally large (RUWE=15.53), possibly indicating a hidden binary companion. Using the Gaia position, proper motion, radial velocity, and parallax of G 73-59, we calculated the 3D galactic velocity ($U_{lsr}$, $V_{lsr}$, $W_{lsr}$), correcting for solar motion using the values of $U_\sun, V_\sun, W_\sun$=(10.0, 5.25, 7.17) \kms\ from \cite{1998MNRAS.298..387D}. The resultant velocities for G 73-59 are $U_{lsr}$, $V_{lsr}$, $W_{lsr}$=(-66.8, -26.5, -39.1) \kms\, placing it as kinematically similar to stars found within the thick disk \citep{2003A&A...410..527B}. No metallicity measurements currently exist in the literature for G 73-59.

CWISE J022737.75+083008.8 (CW0227+0830) is a common proper motion companion with G 73-59. The near infrared spectrum of CW0227+0830 in Figure \ref{fig: 3} classifies it as having a spectral type of L5 (blue). In Figure \ref{fig: 3}, the spectrum of CW0227+0830 is noticeably bluer than typical field dwarfs of the same spectral type. CW0227+0830 has a $J$-$K$ color of 1.29$\pm$0.24 mag, which when compared to the average L5 color of $J$-$K$=1.75$\pm$0.22 mag from \cite{2016ApJS..225...10F}, we see that CW0227+0830 is a blue color outlier, agreeing with its observed spectrum. Looking at CW0227+0830 on the WISE CMD in Figure \ref{fig: 3}, although its location is consistent with typical field L dwarfs, it appears slightly fainter, placing it closer to later-type mid-L dwarfs. The tangential velocity of CW0227+0830, $v_{tan}$=63$\pm$2 \kms, is much more similar to other blue outliers than to typical field dwarfs, which have average tangential velocities of $v_{tan}$=53$\pm$47 \kms\ and $v_{tan}$=26$\pm$19 \kms, respectively \citep{2009AJ....137....1F}, hinting that CW0227+0830 may belong to an older population. We also tested the possibility for the bluer spectral morphology of CW0227+0830 to be due to an unresolved companion, as described in Section \ref{sec:sb}. Table \ref{tab:sb} shows that CW0227+0830 meets the criteria for a weak spectral binary candidate. 

 G 73-59 and CW0227+0830 are separated on the sky by an angular separation of $\sim$16\arcsec. At the distance of G 73-59, that translates to a projected physical separation of $\sim$550 AU. Given the high kinematics of G 73-59, as well as the blue spectral morphology of CW0227+0830, we believe G 73-59 + CW0227+0830 to be an older widely separated M3+L5 (blue) pair.

\subsection{LAMOST J0626+5933AB + CWISE J062648.96+594129.2} 
LAMOST J062631.15+593341.3 and LAMOST J062631.34+593343.6 (hereafter referred to as LAMOST J0626+5933A and LAMOST J0626+5933B) form a close binary pair, with an angular separation of 2.8\arcsec\ and spectral types M1 and M2 respectively \citep{2019ApJS..245...34X}. Only LAMOST J0626+5933B has a corresponding entry in Gaia, so looking at only LAMOST J0626+5933B on the Gaia CMD in Figure \ref{fig: 3}, we see that it appears to be slightly fainter than stars of a similar color. \cite{2019ApJS..245...34X} report multiple metallicities for both components, but only [Fe/H]=-0.88$\pm$0.37 for LAMOST J0626+5933B is listed as reliable. \cite{2022ApJS..260...45D} report metallicities for LAMOST J0626+5933A and LAMOST J0626+5933B (with repeated measurements) of [Fe/H]=-0.45$\pm$0.26 and [Fe/H]=-0.43$\pm$0.26 respectively. Additionally, LAMOST J0626+5933B has a global metallicity measurement of [M/H]=$-0.81^{+0.14}_{-0.20}$ from the Gaia RVS spectrum, as well as an $\alpha$ abundance of [$\alpha$/Fe]=$0.29^{+0.07}_{-0.06}$. There are no reported UV or X-ray detections in the literature, and no TESS light curve is available. We calculated the 3D velocity of LAMOST J0626+5933B as described in Section \ref{sec:0227}, and obtain the values $U_{lsr}$, $V_{lsr}$, $W_{lsr}$=(-85.2, 6.9, -10.4) \kms\, placing it as kinematically similar to stars within the thick disk \citep{2003A&A...410..527B}.

CWISE J062648.96+594129.2 (CW0626+5941) is a common proper motion companion with LAMOST J0626+5933AB. The near infrared spectrum of CW0626+5941, shown in Figure \ref{fig: 3}, classifies it as a spectral type of L2 (blue), as its spectrum is slightly bluer than typical field dwarfs of the same spectral type. CW0626+5941 has a $J$-$K$ color of 1.45$\pm$0.19 mag, slightly bluer than the average L2 color of $J$-$K$=1.51$\pm$0.21 mag \citep{2016ApJS..225...10F}, however it is not blue enough to classify it as a photometric outlier. The tangential velocity of CW0626+59441, $v_{tan}$=66$\pm$4 \kms, is kinematically similar to the average $v_{tan}$=53$\pm$47 \kms\ of known blue outliers \citep{2009AJ....137....1F}.

LAMOST J0626+5933AB and CW0626+5941 are separated on the sky by an angular separation of $\sim$488\arcsec. At the Gaia distance of LAMOST J0626+5933B, this translates to a projected physical separation of $\sim$27,000 AU. Given the reported sub-solar metallicities and large 3D velocity for LAMOST J0626+5933AB, as well as the slightly blue spectral morphology of CW0626+5941 and its $v_{tan}$ similar to blue outliers, we believe LAMOST J0626+5933AB + CW0626+5941 to be a widely separated M1+M2+L2 (blue) triple.

\subsection{WISE J062727.34$-$002826.8 + CWISE J062725.95$-$002843.8}
CWISE J062727.34$-$002826.8 (CW0627-0028A) is classified as a T0 (blue) dwarf via comparison of its near infrared spectrum with the near infrared spectral standards, shown in Figure \ref{fig: 3}. The spectrum of CW0627$-$0028A in Figure \ref{fig: 3} is much bluer, however, than typical field dwarfs of the same spectral type. The $J$-$K$ color of CW0627$-$0028A (1.02$\pm$0.32 mag) is much bluer than the average T0 color of $J$-$K$=1.63$\pm$0.40 mag \citep{2009AJ....137....1F}, agreeing with the bluer spectral morphology of CW0627$-$0028A. The mid infrared color of CW0627$-$0028A, W1-W2=0.36$\pm$0.04 mag, is also bluer than expected, being more consistent with mid-L dwarfs. We investigated the possibility that the bluer colors of CW0627-0028A are due to an unresolved companion, as described in Section \ref{sec:sb}. This analysis (Table \ref{tab:sb}) shows that CW0627-0028A does appear consistent with being a weak spectral binary candidate. Using the best fit spectral type for CW0627$-$0028A of T0, along with a distance estimate from the absolute magnitude - spectral type relations of \cite{2021ApJS..253....7K}, we estimate CW0627$-$0028A to be at distance of 35$\pm$7 pc.

CWISE J062725.95$-$002843.8 (CW0627$-$0028B) is a common proper motion companion with CW0627-0028A. The mid-infrared color of CW0627$-$0028B, W1-W2=0.34$\pm$0.05 mag, is indicative of a mid-L spectral type. However, assuming CW062$-$0028A and CW0627$-$0028B are coeval, CW0627$-$0028B likely also has bluer colors for its spectral type than typical field dwarfs, making an accurate spectral type estimate difficult. As the mid- and near-infrared colors of CW0627$-$0028A and CW0627$-$0028B are similar, we estimate CW0627$-$0028B to likely have a spectral type of early-T. Further spectroscopic observations are required to confirm the spectral type of CW0627$-$0028B, and to determine if it displays a bluer spectrum similar to CW0627$-$0028A.

CW0627$-$0028A and CW0627$-$0028B are separated on the sky by an angular separation of $\sim$27\arcsec. While no parallax for either object exists, using the estimated distance of CW0627$-$0028A, this translates to a projected physical separation of $\sim$860 AU. This separation would make the CW0627$-$0028AB system the widest known T+T binary system, exceeding ULAS J020529.62+142114.0, a T1+T3 binary with a projected separation of 71 AU \citep{2013MNRAS.430.1171D}. Additionally, the CW0627$-$0028AB system would be the widest known field age substellar binary, which is currently CWISE J0146-0508AB (L4+L8 (blue) spectral types) at a separation of $\sim$129 AU \citep{2022ApJ...926L..12S}. However, an accurate parallax for the system is required. Given the blue near infrared spectrum of CW0627$-$0028A, as well as the high co-moving probability of the pair, we believe CW0627$-$0028AB to be a widely separated T0 (blue)+(T) binary.

\subsection{LP 677-81 + CWISE J132857.58$-$063747.4}
LP 677-81 is an estimated M3 dwarf using the Gaia photometric relations of \cite{2019AJ....157..231K}. Looking at LP 677-81 on the Gaia CMD in Figure \ref{fig:1}, we see that it is slightly fainter compared to other stars of the same color. No TESS light curve is available for LP 677-81. However, \cite{2020A&A...635A..43R} report a rotational period of $P_{rot}=23\pm23$ days. From looking at LP 677-81 on the color-period diagram shown in Figure \ref{fig: 9}, we find that its age appears to be at least $>$650 Myr. No metallicity measurements have been reported in the literature for LP 677-81, and its Gaia DR3 RUWE of 1.07 indicates that it is likely not an unresolved binary. Using the Gaia kinematics of LP 677-81, we calculated its 3D space motion as described in Section \ref{sec:0227}, finding the velocities $U_{lsr}$, $V_{lsr}$, $W_{lsr}$=(62.9, -51.1, 0.7) \kms\, making LP 677-81 kinematically similar to the thick disk population \citep{2003A&A...410..527B}. .

CWISE J132857.58$-$063747.4 (CW1328$-$0637) is a common proper motion companion with LP 677-81. We classify CW1328$-$0637 as spectral type L3 (blue) based on its near infrared spectrum in Figure \ref{fig: 3}. As can be seen in Figure \ref{fig: 3}, the spectrum of CW1328$-$0637 is much bluer than typical field dwarfs of the same spectral type. The near infrared color of CW1328$-$0627, $J$-$K$=0.96$\pm$0.07 mag, indicates it as a blue color outlier when compared with the average for that spectral type, $J$-$K$=1.61$\pm$0.22 mag \citep{2016ApJS..225...10F}, which is in agreement with the blue morphology of its observed spectrum. Looking at CW1328$-$0637 in the WISE CMD in Figure \ref{fig:1} and adopting the parallax of its suspected host, we see that it appears faint for its spectral type, with an absolute magnitude placing it among mid to late L dwarfs. We find that CW1328-0637 is a weak spectral binary candidate (Table \ref{tab:sb}), however \cite{2014ApJ...794..143B} found that blue L-dwarfs are a major contaminant when investigating spectral binaries. Adopting the Gaia parallax of LP 677-81 we find that CW1328$-$0637 has a tangential velocity of $v_{tan}=77\pm12$ \kms, similar to the tangential velocity of other blue color outliers \citep{2009AJ....137....1F}. 

LP 677-81 and CW1328$-$0637 are separated on the sky by an angular separation of $\sim$19\arcsec. At the distance of LP 677-81, this translates to a projected physical separation of $\sim$1400 AU. Due to LP 677-81 appearing fainter in the Gaia CMD, along with the bluer near infrared spectrum of CW1328$-$0637, we believe LP 677-81 + CW1328$-$0637 to be a (M3)+L3 (blue) widely separated binary. 
 
\subsection{CWISE J160653.16$-$103210.6 + CWISE J160654.19$-$103214.7}
CWISE J160653.16$-$103210.6 (CW1606$-$1032A) is a high proper motion star with an optical classification of M9 based on our Kast optical spectrum. Looking at CW1606$-$1032A on the Gaia CMD in Figure \ref{fig:1}, we see that it is in good agreement with typical field stars of the same color. No information on rotational period, metallicity, H$\alpha$, X-ray, or UV emission is available in the literature. While we are not able to compute the full 3D velocity of CW1606$-$1032A, as no radial velocity measurement exists, using the Gaia distance for CW1606$-$1032A we find that it has a tangential velocity of $v_{tan}$=25 \kms. Comparing this tangential velocity with the average tangential velocities computed in \cite{2009AJ....137....1F}, we see that CW1606$-$1032A has a tangential velocity more similar to field late M dwarfs. The near infrared color of CW1606$-$1032A, $J$-$K$=1.03$\pm$0.06 mag, is also normal for typical field objects of the same spectral type. 

CWISE J160654.19$-$103214.7 (CW1606$-$1032B) is a common proper motion companion with CW1606$-$1032A. Comparing the near infrared spectrum of CW1606$-$1032B in Figure \ref{fig: 3} with the near infrared spectral standards, we classify CW1606$-$1032B as spectral type L7 (blue), as the spectrum of CW1606-1032B is much bluer than typical field dwarfs (Figure \ref{fig: 3}). Additionally, the spectrum of CW1606$-$1032B is suggestive of stronger FeH absorption bands, a feature observed in objects with sub-solar metallicity \citep{2003ApJ...592.1186B,2010ApJS..190..100K}. Looking at CW1606$-$1032B on the WISE CMD using the parallax of its suspected host, we see that it lies amongst other late-type L dwarfs. The near infrared color of CW1606$-$1032B, $J$-$K$=1.58$\pm$0.05 mag, while bluer than average field objects of the same spectral type, is not enough to classify it as a blue outlier. We find that CW1606$-$1032B is a weak spectral binary candidate (Table \ref{tab:sb}) following the methods described in Section \ref{sec:sb}, however as noted in \cite{2014ApJ...794..143B}, blue L-dwarfs are a typical spectral binary contaminant. 

CW1606$-$1032A and CW1606$-$1032B are separated on the sky by an angular separation of $\sim$16\arcsec. At the Gaia distance of CW1606$-$1032A, this translates into a projected physical separation of $\sim$640 AU. While the near infrared spectrum of CW1606$-$1032B resembles the morphology observed in some sub-solar metallicity objects, the kinematics and colors of CW1606$-$1032A are typical for field age dwarfs. We therefore suggest that CW1606$-$1032AB are a field-age M9+L7 (blue) widely separated binary. 

\subsection{SRC J2029$-$7910 + CWISE J202934.80$-$791013.1}
SRC J2029$-$7910 is a high proper motion star which we classify as spectral type M6 (blue) based on its near infrared spectrum. The spectrum of SCR J2029$-$7910 in Figure \ref{fig: 5} however has a much bluer appearance than the spectral standard. Looking at SCR J2029$-$7910 on the Gaia CMD in Figure \ref{fig:1}, we see that it appears fainter compared to objects of the same spectral type. The Gaia RUWE of SCR J2029$-$7910 is anomalously high (RUWE=1.88), hinting at the possible existence of a hidden companion. Additionally, SCR J2029$-$7910 has an entry in the Gaia DR3 non-single stars table, which flags SCR J2029$-$7910 as a likely binary through the analysis of its astrometric acceleration. Using the Gaia DR3 MSC model (which treats sources as though they were an unresolved binary), SCR J2029$-$1032 has a sub-solar metallicity of [M/H]$\approx\-0.20$. No information on rotation period, X-Ray, or UV emission is available in the literature. Using the full Gaia kinematics for SCR J2029$-$7910, we calculated its 3D space motions as described in Section \ref{sec:0227}, obtaining the velocities $U_{lsr}$, $V_{lsr}$, $W_{lsr}$=(-34.8, -37.6, -29.2) \kms. This 3D velocity places SCR J2029-7910 as more kinematically consistent with the thin disk population \citep{2003A&A...410..527B}.

CWISE J202934.80$-$791013.1 (CW2029$-$7910) is a common proper motion companion with SCR J2029$-$7910. CW2029$-$7910 was also identified as being statistically the same distance as SCR J2029$-$7910, with parallaxes of $\varpi$=19.63$\pm$0.66 mas and $\varpi$=20.12$\pm$0.04 mas, respectively. \cite{2018AA...619L...8R} assigned CW2029-7910 a phototype of L0.5 young. However we classify CW2029$-$7910 as spectral type L1 (blue) based on its near infrared spectrum in Figure \ref{fig: 3}. Compared to the near infrared spectral standard as shown in Figure \ref{fig: 3}, the spectrum of CW2029$-$7910 is much bluer, and is suggestive of stronger FeH bands and K I lines, signatures of lower metallicity. Looking at CW2029$-$7910 on the WISE CMD in Figure \ref{fig:1}, we find that it appears slightly faint for its spectral type. On the Gaia CMD in Figure \ref{fig:1} however, CW2029$-$7910 appears near normal compared to field objects of the same color. CW2029-7910 receives gravity classification of FLD-G (Table \ref{table: 4}) from the indices of \cite{2013ApJ...772...79A}. No metallicity, rotational period, X-Ray, or UV emission data are available in the literature. CW2029$-$7910 has a near infrared color of $J$-$K$=1.02$\pm$0.17 mag, placing it as a borderline blue color outlier. Table \ref{tab:sb} shows that CW2029$-$7910 is a strong spectral binary candidate. The tangential velocity of CW2029-7910 is $v_{tan}$=71.5 \kms, similar to other color outliers in \citep{2009AJ....137....1F}, suggestive that it may be an older dwarf star. 

SCR J2029$-$7910 and CW2029$-$7910 are separated on the sky by an angular separation of $\sim$34\arcsec. At the Gaia distance of SCR J2029$-$7910, this translates to a projected physical separation of $\sim$1700 AU. Given the bluer spectral morphologies of SCR J2029$-$7910 and CW2029$-$7910, as well as the sub-solar metallicity of SCR J2029$-$7910, we believe that SCR J2029$-$7910 + CW2029$-$7910 are a widely separated low-metallicity M6 (blue)+L1 (blue) system.

\section{Unusually Red Spectra}
There exist some brown dwarfs that display red near-infrared colors, yet show no indications of youth. It has been suggested that an excess of dust, consisting of micron sized grains, high in the brown dwarf atmospheres would preferentially suppress flux at shorter wavelengths, resulting in the observed reddening \citep{2008ApJ...678.1372C,2014MNRAS.439..372M,2016ApJ...830...96H}. Viewing angle has also been shown to be correlated with color anomalies, where objects viewed equator-on have been observed to be redder than other objects viewed at different latitudes \citep{2017ApJ...842...78V}. This correlation has been explained by cloud properties, as equatorial latitudes are cloudier than polar latitudes \citep{2023ApJ...954L...6S}. Below we discuss the objects in our sample which have redder than expected near-infrared spectra which cannot be attributed to youth.

\subsection{LSPM J0738+5254 + CWISE J073831.31+525453.7}
LSPM J0738+5254 is classified as spectral type M3 \citep{2019yCat.5164....0L}. On the Gaia CMD in Figure \ref{fig:1}, LSPM J0738+5254 appears normal for its spectral type. \cite{2022ApJS..260...45D} find a near solar metallicity of [Fe/H]=$0.12\pm0.26$, whereas \cite{2020ApJS..246....9Z} find a metallicity of [M/H]=$-0.62\pm0.13$. \cite{2019ApJS..243...28L} find an H$\alpha$ equivalent width of 0.236$\pm$0.021 \AA, placing it in the inactive region of the H$\alpha$--age relations of \cite{2021AJ....161..277K}, making an accurate age determination difficult. \cite{2021AJ....161..277K} point out that while stars are found in the inactive region at all ages, the large majority of inactive stars tend to be those with older ages.  While we can not resolve the metallicity variation noted above, the inactivity favors the lower metallicity and older age. No light curve is available from TESS or in the literature. Running the full Gaia kinematics of LSPM J0738+5254 through BANYAN $\Sigma$, we find that it receives a 99.9\% field probability.

CWISE J073831.31+525453.7 (CW0738+5254) is a common proper motion companion with LSPM J0738+5254. While the $J$-band peak of CW0738+5254's spectrum is most similar to that of the L4 near infrared standard as shown in Figure \ref{fig: 3}, its continuum appears too dissimilar from the standard to type it through comparison of the J$-$band alone. A band-by-band classification (cf. \citealt{2014MNRAS.439..372M,2018AJ....155...34C}) shows that CW0738+5254 is indeed best fit by the L4 near infrared spectral standard. However, its spectrum is much redder than typical field dwarfs of the same spectral type. Using the indices of \cite{2013ApJ...772...79A} we find that CW0738+5254 receives a gravity score of FLD-G (Table \ref{table: 4}, indicating an object with field age surface gravity. No other features of youth, such as the distinctive triangular $H$-band, are visible in CW0738+5254's spectrum in Figure \ref{fig: 3}. Looking at its position on the WISE CMD (Figure \ref{fig:1}), using the parallax of its suspected host, we find that CW0738+5254 appears near normal for its spectral type, although slightly fainter. Using an estimated parallax for CW0738+5254 based on spectral type - absolute magnitude estimates, we ran the kinematics of CW0738+5254 through BANYAN $\Sigma$, finding a 5.7\% membership probability for the AB Doradus moving group and a 94.3\% field probability. 

LSPM J0738+5254 and CW0738+5254 are separated on the sky by an angular separation of $\sim$11\arcsec. At the Gaia distance of LSPM J0738+5254, this translates to a projected physical separation of $\sim$530 AU. As we do not find any distinguishing characteristics of youth for either CW0738+5254 or LSPM J0738+5254, we believe LSPM J0738+5254 + CW0738+5254 to be a field age widely separated M3+L4 (red) binary.

\subsection{2MASS J09435055+3356550 + CWISE J094352.22+335639.1}
2MASS J09435055+3356550 (2M0943+3356) is a high proper motion star with an estimated phototype of M3.5 \citep{2016MNRAS.457.2192C}. Looking at 2M0943+3356 on the Gaia CMD in Figure \ref{fig:1}, we see that it appears near normal for its spectral type. \cite{2022AJ....163..152S} finds a near solar metallicity of [Fe/H]=0.067$\pm$0.007. No other information, such as rotational period, H$\alpha$, UV, or X-Ray emission is available in the literature. Running the full Gaia kinematics of 2M0943+3356 through BANYAN $\Sigma$, we find a 99.9\% field probability.

CWISE J094352.22+335639.1 (CW0943+3356) is a common proper motion companion with 2M0943+3356. The near infrared spectrum of CW0943+3356 in Figure \ref{fig: 3} is matched closest with the L2 near infrared spectral standard. However, the spectrum of CW0943+3356 in Figure \ref{fig: 3} appears to have more flux at longer wavelengths, giving it a redder color than the spectral standard. Using the indices of \cite{2013ApJ...772...79A} we find that CW0943+3356 receives a gravity score of FLD-G (Table \ref{table: 4}). Looking at CW0943+3356 on the WISE CMD in Figure \ref{fig:1}, using the parallax of its suspected host, we see that it appears fainter than expected for its spectral type, instead occupying a region typically populated by mid-L type dwarfs. Running the proper motions and estimated distance through BANYAN $\Sigma$, we find a 0.7\% membership probability for the Argus moving group and a 99.3\% field probability.

2M0943+3356 and CW0943+3356 are separated on the sky by an angular separation of $\sim$28\arcsec. At the Gaia distance of 2M0943+3356, this translates to a projected physical separation $\sim$1,820 AU. With kinematics for both objects, as well as the gravity sensitive spectral indices of CW0943+3356, consistent with the field population, we do not believe the reddened spectrum of CW0943+3356 is due to youth. We therefore classify 2M0943+3356 + CW0943+3356 as a widely separated (M3.5)+L2 (red) system.

\subsection{2MASS J13032992+5127582 + CWISE J130329.90+512754.0}
\begin{figure*}
    \centering
    \includegraphics[width=1\textwidth]{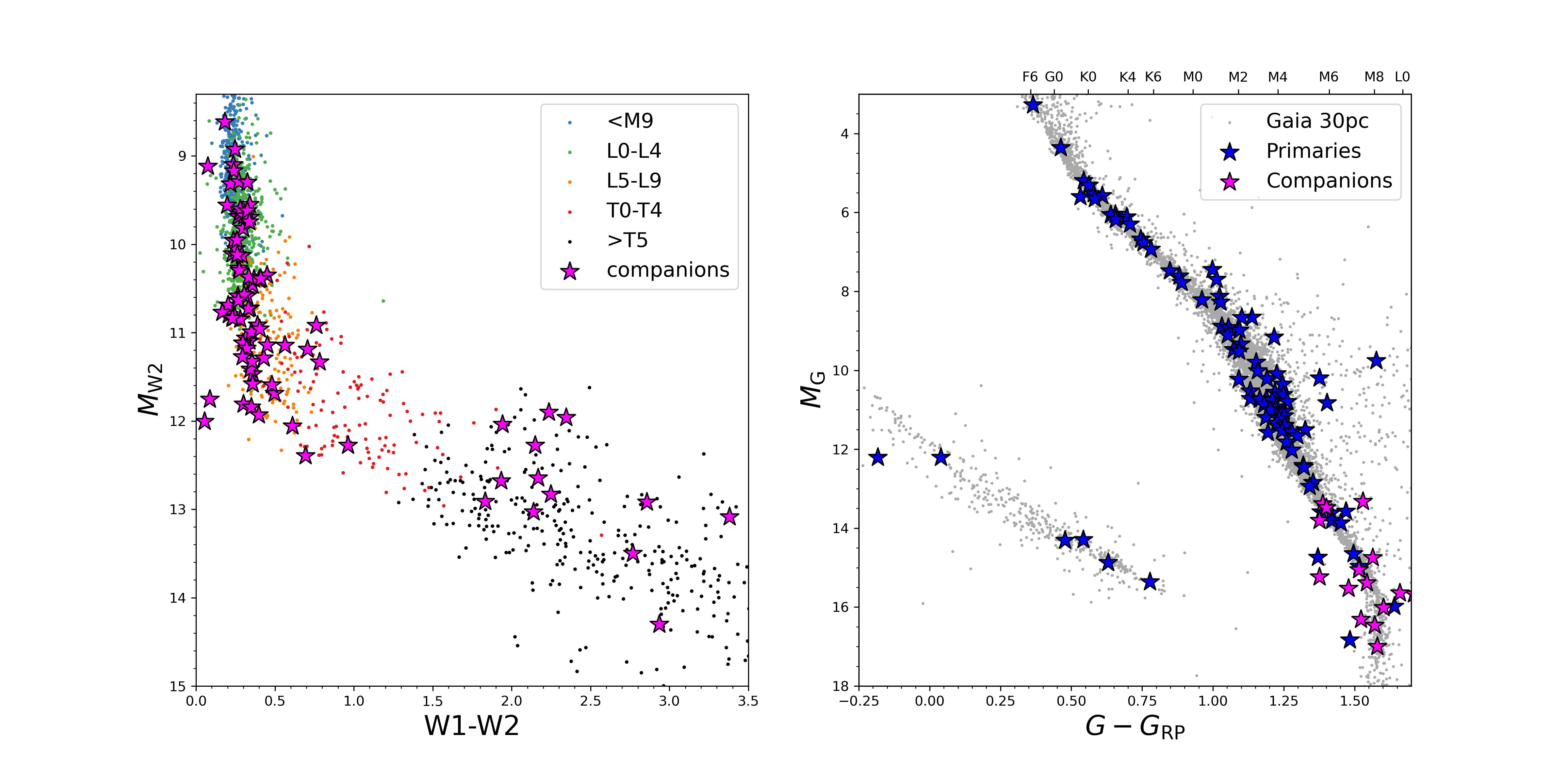}
    \caption{\textbf{Left:} CMD of potential companion objects using WISE photometry, adopting their host's parallax if necessary. Plotted with the sample are known ultracool dwarfs with parallaxes taken from The UltracoolSheet \citep{best_william_m_j_2020_4169085}, color coded by spectral type. \textbf{Right:} CMD of potential host objects (as blue stars) using Gaia DR3 photometry. Any suspected companions with Gaia DR3 parallaxes and astrometry are plotted as pink stars. Plotted for reference is the Gaia DR3 30 pc sample.}
    \label{fig:1}
\end{figure*}

2MASS J13032992+5127582 (2M1303+5127) is phototyped as spectral type M8 \citep{2018A&A...619L...8R}. Looking at 2M1303+5127 on the Gaia CMD in Figure \ref{fig:1}, we see that it appears near normal for its spectral type. No additional information is available in the literature. Running the Gaia proper motions and parallax for 2M1303+5127 through BANYAN $\Sigma$, we find a 20.4\% membership probability for the Carina-Near moving group and a 79.6\% field probability. 

\begin{figure*}
    \centering
    \includegraphics[scale=0.7]{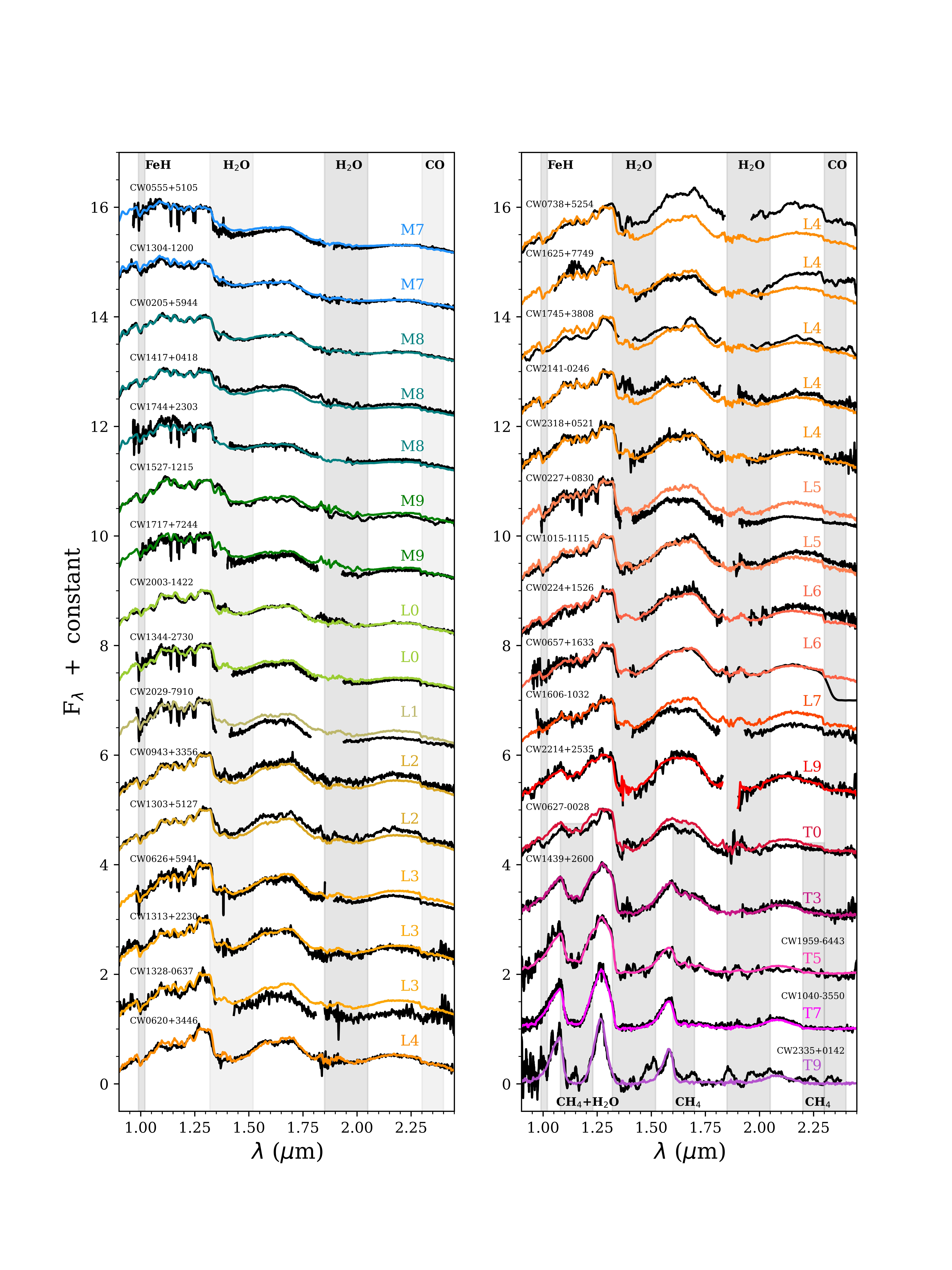}
    \caption{Near-Infrared spectra of the companion components plotted in black. Over-plotted in color are the spectral standards as described in Section \ref{sec:analysis}. All spectra are normalized between 1.27-1.29 $\mu$m and separated by a constant. \\
    \textbf{Spectra obtained with}: SpeX-- CW0205+5944, CW0224+1526, CW0620+3446, CW0627-0028, CW0738+5254, CW0943+3356, CW1015-1115, CW1037-0507, CW1303+5127, CW1304-1200, CW1313+2230, CW1328-0637, CW1417+0418, CW1439+2600, CW1527-1215, CW1745+3807, CW2003-1422, CW2141-0246, CW2214-0246, CW2318+0521;
    NIRES-- CW0227+0830, CW0555+5105, CW0626+5941, CW1625+7749, CW1717+7244;
    FIRE-- CW1040-3550, CW1959-6443, CW2335+0142;
    Triplespec-- CW1606; NIHTS--CW0657+1633; ArcoIRIS--CW2029-7910.}
    \label{fig: 3}
\end{figure*}

CWISE J130329.90+512754.0 (CW1303+5127) is a common proper motion companion with 2M1303+5127. While CW1303+5127 does have a Gaia DR3 source ID, there is no available astrometry for this source. The near infrared spectrum of CW1303+5127 in Figure \ref{fig: 3} is best fit by the L2 near infrared spectral standard. CW1303+5127's spectrum in Figure \ref{fig:1} appears redder than the spectral standard however, as can be seen in Figure \ref{fig: 3}. Using the indices of \cite{2013ApJ...772...79A} we find that CW1303+5127 receives a gravity score of FLD-G (Table \ref{table: 4}. Looking at CW1303+5127 on the WISE CMD (Figure \ref{fig:1}) using the parallax of its suspected host, we see that it appears near normal for its spectral type. Running the proper motions and estimated distance of CW1303+5127 through BANYAN $\Sigma$, we find a 4.1\% probability of membership to the Carina-Near moving group, a 59.9\% probability of membership to the Argus moving group, and a 36.0\% field probability.  

2M1303+5127 and CW1303+5127 are separated on the sky by an angular separation of $\sim$5\arcsec. At the Gaia distance of 2M1303+5127, this translates to a projected physical separation of $\sim$240 AU. While both CW1303+5127 and 2M1303+5127 appear to receive an elevated probabilities of belonging to the Argus or Carina-Near moving groups, the gravity score of FLD-G for CW1303+5127 suggests the possibility of the pair being field age contaminants. Higher accuracy proper motions for CW1303+5127, as well as radial velocities for both components, are required in order to confirm any potential memberships to young moving groups. As the age of this system is ambiguous at this time, we believe 2M1303+5127 + CW1303+5127 to be a widely separated field age (M8)+L2 (red).

\startlongtable
\begin{longrotatetable}
\begin{deluxetable*}{lcccccccccccc}
\tabletypesize{\tiny}
\tablewidth{0pt}
\tablenum{1}
\tablecolumns{12}
\tablecaption{High Probability Co-moving Systems.\label{table: 3}}
\tablehead{ 
\colhead{Name} & 
\colhead{SpType$^a$} &
\colhead{Sp$\_$ref} &
\colhead{$\mu_{\alpha}$} & 
\colhead{$\mu_{\delta}$} & 
\colhead{RV} & 
\colhead{vsin(i)} &
\colhead{Plx} &
\colhead{Ast$\_$ref} &
\colhead{Sep} &
\colhead{Sep} &
\colhead{Probability} \\
 & & & \colhead{(mas/yr)} & \colhead{(mas/yr)} & \colhead{\kms} & \colhead{\kms} & \colhead{(mas)} & & \colhead{$\arcsec$} & \colhead{(au)} & \colhead{$\%$}}
\startdata
TYC 8025-428-1 (0000-4812)	&	G5	&	8	&	166.67	$\pm$	0.01	&	-54.58	$\pm$	0.01	&	2.84	$\pm$	0.16	&	6.85	$\pm$	3.05	&	11.42	$\pm$	0.01	&	2	&	---	&	---	&	100	\\	
CWISE J000021.45-481314.9	&	(L4)	&	1	&	172.91	$\pm$	20.50	&	-67.03	$\pm$	19.20	&		---		&		---		&		---		&	3	&	57	&	5,000	&		\\	\hline
WDJ002027.83-153546.50	&	(WD)	&	9	&	173.66	$\pm$	0.30	&	6.55	$\pm$	6.64	&		$\pm$		&		---		&	10.52	$\pm$	0.32	&	2	&	---	&	---	&	100	\\	
CWISE J002029.72-153527.5	&	(M9)	&	1	&	173.03	$\pm$	13.10	&	16.64	$\pm$	12.50	&		---		&		---		&		---		&	3	&	31	&	3,000	&		\\	\hline
UCAC3 113-933 (0021-3346)	&	(M0)	&	1	&	113.50	$\pm$	0.02	&	-51.36	$\pm$	0.03	&	12.47	$\pm$	0.52	&	11.22	$\pm$	5.42	&	19.02	$\pm$	0.02	&	2	&	---	&	---	&	97.7	\\	
CWISE J002101.45-334631.5	&	(T1)	&	1	&	166.95	$\pm$	60.20	&	84.15	$\pm$	56.60	&		---		&		---		&		---		&	3	&	116	&	6,100	&		\\	\hline
2MASS J00215427-4514590	&	(M5)	&	1	&	201.94	$\pm$	0.02	&	-99.17	$\pm$	0.03	&	11.06	$\pm$	4.26	&		---		&	18.46	$\pm$	0.04	&	2	&	---	&	---	&	99.7	\\	
CWISE J002159.02-451434.4	&	(T3)	&	1	&	178.56	$\pm$	68.50	&	-75.73	$\pm$	68.80	&		---		&		---		&		---		&	3	&	54	&	2,900	&		\\
\enddata
\tablecomments{$^a$ Spectral types in parenthesis, e.g. (SpT), denote estimated spectral types. \\ 
$^b$ Parallax measurement from \cite{2021ApJS..253....7K}. \\
References: 1-This work; 2-\cite{gaia2023}; 3-\cite{2021ApJS..253....8M}; 4-\cite{2023AJ....165...36M}; 5-\cite{2021AJ....161..192N}; 6-\cite{2020ApJ...889...74M}; 7-\cite{2016arXiv161205560C}; 8-\cite{1939cfsc.book.....S}; 9-\cite{2021MNRAS.508.3877G}; 10-\cite{2010PASP..122.1437P}; 11-\cite{2019ApJS..245...34X}; 12-\cite{1993BICDS..43....5T}; 13-\cite{2016MNRAS.457.2192C}; 14-\cite{1978AJ.....83.1090P}; 15-\cite{2006AA...460..695T}; 16-\cite{1993yCat.3135....0C}; 17-\cite{2014MNRAS.443.2561G}; 18-\cite{2021AJ....161...63W}; 19-\cite{1986AJ.....91..144S}; 20-\cite{2011AA...530A.114K}; 21-\cite{2017MNRAS.472.4173R}; 22-\cite{1988AAS...74..449R}; 23-\cite{2016AA...589A..49S}; 24-\cite{2021AA...645A.100S}; 25-\cite{2018AA...619L...8R}; 26-\cite{2006AJ....132..161G}; 27-\cite{2015MNRAS.449.3651M}; 28-\cite{1958PMcCO..13b...0V}; 29-\cite{1975mcts.book.....H}; 30-\cite{2016AA...587A..51S}; 31-\cite{2021ApJS..253...19Z}; 32-\cite{2009ApJ...699..649S}; 33-\cite{2007AstBu..62..117M} }
\end{deluxetable*} 
\end{longrotatetable}

\startlongtable
\begin{longrotatetable}
\begin{deluxetable*}{lcccccccccccc}
\tabletypesize{\tiny}
\tablewidth{0pt}
\tablenum{2}
\tablecolumns{10}
\tablecaption{Catalog Photometry. \label{table: 2}}
\tablehead{
\colhead{Name} &
\colhead{$G_{G}^a$} &
\colhead{$G_{BP}^b$} &
\colhead{$G_{RP}^c$} &
\colhead{J$^d$} &
\colhead{H$^e$} &
\colhead{K$^f$} &
\colhead{NIR ref.} & 
\colhead{W1$^g$} & 
\colhead{W2$^h$} \\
 &
\colhead{(mag)} &
\colhead{(mag)} &
\colhead{(mag)} &
\colhead{(mag)} &
\colhead{(mag)} &
\colhead{(mag)} &
 &
\colhead{(mag)} &
\colhead{(mag)}
}
\startdata
TYC 8025-428-1 (0000-4812)	&	10.235	$\pm$	0.003	&	10.649	$\pm$	0.003	&	9.656	$\pm$	0.004	&	9.02	$\pm$	0.03	&	8.59	$\pm$	0.03	&	8.58	$\pm$	0.02	&	1	&	8.51	$\pm$	0.01	&	8.56	$\pm$	0.01	\\	
CWISE J000021.45-481314.9	&	---			&	---			&	---			&	17.89	$\pm$	0.04	&	17.00	$\pm$	0.04	&	16.37	$\pm$	0.04	&	3	&	15.72	$\pm$	0.02	&	15.51	$\pm$	0.04	\\	\hline
WDJ002027.83-153546.50	&	19.199	$\pm$	0.004	&	19.557	$\pm$	0.031	&	18.721	$\pm$	0.027	&		---		&		---		&		---		&	---	&	18.22	$\pm$	0.16	&	18.25	$\pm$	0.54	\\	
CWISE J002029.72-153527.5	&	20.962	$\pm$	0.017	&	21.330	$\pm$	0.245	&	19.332	$\pm$	0.082	&	16.34	$\pm$	0.11	&	15.74	$\pm$	0.13	&	15.16	$\pm$	0.13	&	1	&	14.94	$\pm$	0.02	&	14.64	$\pm$	0.02	\\	\hline
UCAC3 113-933 (0021-3346)	&	11.820	$\pm$	0.003	&	12.756	$\pm$	0.003	&	10.860	$\pm$	0.004	&	9.73	$\pm$	0.02	&	9.02	$\pm$	0.02	&	8.87	$\pm$	0.02	&	1	&	8.76	$\pm$	0.01	&	8.73	$\pm$	0.01	\\	
CWISE J002101.45-334631.5	&	---			&	---			&	---			&	19.94	$\pm$	0.10	&	19.63	$\pm$	0.15	&		---		&	8	&	17.91	$\pm$	0.11	&	15.56	$\pm$	0.05	\\	\hline
2MASS J00215427-4514590	&	15.483	$\pm$	0.003	&	17.226	$\pm$	0.007	&	14.224	$\pm$	0.004	&	12.46	$\pm$	0.02	&	11.85	$\pm$	0.02	&	11.57	$\pm$	0.02	&	1	&	11.41	$\pm$	0.01	&	11.22	$\pm$	0.01	\\	
CWISE J002159.02-451434.4	&	---			&	---			&	---			&		---		&		---		&		---		&	---	&	18.09	$\pm$	0.12	&	15.94	$\pm$	0.05	\\
\enddata
\tablecomments{References: 1- 2MASS \citep{2006AJ....131.1163S}; 2- 2MASS reject; 3- VHS \citep{2012AA...548A.119C}; 4- UHS \citep{2018MNRAS.473.5113D}; 5-UKIDSS LAS \citep{2007MNRAS.379.1599L}; 6- UKIDSS GPS \citep{2007MNRAS.379.1599L}; 8- UKIDSS GCS \citep{2007MNRAS.379.1599L}; 9-\cite{2020ApJ...889...74M}; 10-\cite{2015ApJ...814..118B}; 11-\cite{2017ApJ...842..118L}\\
\tablenotetext{a}{Gaia DR3 G magnitude.}
\tablenotetext{b}{Gaia DR3 BP magnitude.}
\tablenotetext{c}{Gaia DR3 RP magnitude.} 
\tablenotetext{d}{J-Band magnitude, both MKO and 2MASS filters.}
\tablenotetext{e}{H-Band magnitude.}
\tablenotetext{f}{K-Band (and Ks) magnitude, both MKO and 2MASS filters.}
\tablenotetext{g}{CatWISE2020 W1 magnitude.}
\tablenotetext{h}{CatWISE2020 W2 magnitude.}}
\end{deluxetable*} 
\end{longrotatetable}

\startlongtable
\begin{longrotatetable}
\begin{deluxetable*}{lcccccccccccc}
\tabletypesize{\tiny}
\tablewidth{0pt}
\tablenum{3}
\tablecolumns{12}
\tablecaption{Low Probability Co-moving Systems.\label{table: 10}}
\tablehead{ 
\colhead{Name} & 
\colhead{SpType$^a$} &
\colhead{Sp$\_$ref} &
\colhead{$\mu_{\alpha}$ (mas/yr)} & 
\colhead{$\mu_{\delta}$ (mas/yr)} & 
\colhead{RV (\kms)} & 
\colhead{vsin(i) (\kms)} &
\colhead{Plx (mas)} &
\colhead{Ast$\_$ref} & 
\colhead{Sep ($\arcsec$)} &
\colhead{Sep (au)} &
\colhead{Probability ($\%$)} }
\startdata
LP 293-22	&	(M4)	&	10	&	227.42	$\pm$	0.12	&	-77.48	$\pm$	0.09	&		---		&		---		&	7.17	$\pm$	0.11	&	2	&	---	&	---	&	0	\\	
J002820.38+320128.9	&	(L5)	&	1	&	214.49	$\pm$	5.24	&	-120.90	$\pm$	5.24	&		---		&		---		&		---		&	7	&	580	&	81,000	&		\\	\hline
G 1-44	&	(M4)	&	11	&	383.49	$\pm$	0.04	&	39.93	$\pm$	0.03	&	-20.64	$\pm$	0.34	&	7.75	$\pm$	4.25	&	34.99	$\pm$	0.04	&	2	&	---	&	---	&	88.6	\\	
2MASS J01020186+0355405	&	L9	&	8	&	355.90	$\pm$	9.50	&	84.35	$\pm$	9.20	&		---		&		---		&		---		&	3	&	590	&	17,000	&		\\	\hline
LSPM J0230+2400	&	(M4.5)	&	10	&	167.60	$\pm$	0.10	&	-188.50	$\pm$	0.09	&		---		&		---		&	17.53	$\pm$	0.09	&	2	&	---	&	---	&	73.1	\\	
CWISE J023039.27+240502.8	&	L4	&	1	&	246.44	$\pm$	49.92	&	-83.72	$\pm$	51.11	&		---		&		---		&		---		&	5	&	302	&	17,000	&		\\	\hline
Gaia EDR3 469538198602403200	&	(M8.5)	&	12	&	99.32	$\pm$	0.48	&	-163.89	$\pm$	0.46	&		---		&		---		&	13.09	$\pm$	0.46	&	2	&	---	&	---	&	0	\\	
CWISE J035823.50+562453.2	&	L7	&	1	&	9.69	$\pm$	8.60	&	-53.60	$\pm$	8.40	&		---		&		---		&		---		&	3	&	20	&	1,500	&		\\
\enddata
\tablecomments{Summary of candidate systems which received below the 90$\%$ threshold. \\
$^a$ Spectral types given within parenthesis are estimates.
References: 1-This work; 2-\cite{gaia2023}; 3-\cite{2021ApJS..253....8M}; 4-\cite{2023AJ....165...36M}; 5-\cite{2021AJ....161..192N}; 6-\cite{2020ApJ...889...74M}; 7-\cite{2016arXiv161205560C}; 8-\cite{2016ApJ...817..112S}; 9-\cite{2015ApJ...814..118B}; 10-\cite{2016MNRAS.457.2192C}; 11-\cite{1984AJ.....89.1229R}; 12-\cite{2018AA...619L...8R}; 13-\cite{2019MNRAS.489.5301C}; 14-\cite{2015RAA....15.1154Z}; 15-\cite{2021ApJS..253...19Z}; 16-\cite{2021MNRAS.508.3877G}; 17-\cite{2014MNRAS.443.2561G}; 18-\cite{2019ApJ...877...60B}; 19-\cite{1975mcts.book.....H}; 20-\cite{2006AJ....132..161G}}
\end{deluxetable*} 
\end{longrotatetable}

\begin{deluxetable*}{lcc}
\tabletypesize{\tiny}
\tablewidth{0pt}
\tablenum{4}
\tablecolumns{4}
\tablecaption{Spectroscopic Observations.\label{table: 1}}
\tablehead{
\colhead{Name$^a$} &
\colhead{Telescope/Instrument} &
\colhead{UT Date Observed}}
\startdata
 & & \\
Secondary Components & & \\
\hline
CWISE J020538.20+594452.2 & IRTF/SpeX & January 8, 2022\\
CWISE J022737.75+083008.8 & Keck/NIRES & October 27, 2018\\
CWISE J023039.27+240502.8 & IRTF/SpeX & December 23, 2020\\
CWISE J035823.50+562453.2 & IRTF/SpeX & December 23, 2020\\
CWISE J055515.83+510514.0 & Keck/NIRES & February 24, 2021\\
CWISE J062000.01+344641.3 & IRTF/SpeX & November 23, 2020\\
\enddata
\tablecomments{$^a$Names which do not contain coordinate information within them are listed with abbreviated coordinates in parenthesis.}
\end{deluxetable*}

\begin{deluxetable*}{lccccccc}
\tabletypesize{\scriptsize}
\tablewidth{0pt}
\tablenum{5}
\tablecolumns{8}
\tablecaption{Gravity Indices \label{table: 4}}
\tablehead{
\colhead{Name} &
\colhead{Sptype} &
\colhead{FeH$_Z$} &
\colhead{VO$_Z$} &
\colhead{KI$_J$} &
\colhead{H-Cont} &
\colhead{Scores} &
\colhead{Class}
}
\startdata
CWISE 0205+5944	&	M8	&	1.20	$\pm$	0.01	&	1.03	$\pm$	0.00	&	1.03	$\pm$	0.00	&	0.98	$\pm$	0.00	&	0	---	2	1	&	INT-G	\\
CWISE 0227+0830	&	L5 (blue)	&	1.77	$\pm$	0.64	&	0.98	$\pm$	0.02	&	1.17	$\pm$	0.01	&	0.80	$\pm$	0.01	&	0	---	0	0	&	FLD-G	\\
CWISE 0224+1526A&	L6 (red)	&	1.12	$\pm$	0.05	&	0.96	$\pm$	0.03	&	1.04	$\pm$	0.02	&	0.88	$\pm$	0.02	&	---	---	---	1	&	INT-G	\\
CWISE 0555+5105	&	M7	&	1.17	$\pm$	0.01	&	1.01	$\pm$	0.00	&	1.08	$\pm$	0.00	&	0.99	$\pm$	0.00	&	0	---	0	2	&	FLD-G	\\
CWISE 0620+3446	&	L4	&	1.24	$\pm$	0.02	&	1.01	$\pm$	0.01	&	1.07	$\pm$	0.01	&	0.81	$\pm$	0.01	&	1	0	2	0	&	FLD-G	\\
CWISE 0626+5941	&	L3 (blue)	&	1.51	$\pm$	0.18	&	1.05	$\pm$	0.01	&	1.18	$\pm$	0.01	&	0.86	$\pm$	0.01	&	0	0	0	0	&	FLD-G	\\
CWISE 0657+1653	&	L6	&	1.36	$\pm$	0.12	&	0.89	$\pm$	0.04	&	1.09	$\pm$	0.02	&	0.88	$\pm$	0.01	&	--	--	--	1	&	INT-G	\\
CWISE 0738+5254	&	L4 (red)	&	1.17	$\pm$	0.10	&	0.89	$\pm$	0.05	&	0.91	$\pm$	0.04	&	0.83	$\pm$	0.03	&	1	0	2	0	&	FLD-G	\\
\enddata
\tablecomments{Gravity-sensitive indices from \cite{2013ApJ...772...79A}. Gravity scores for each index is provided in the penultimate column, with dashes indicating the value is out of range. The resulting gravity classes are provided in the final column.}
\end{deluxetable*}

\begin{longrotatetable}
    \begin{deluxetable*}{lccccccccccccccc}
        \tabletypesize{\tiny}
        \tablewidth{0pt}
        \tablenum{6}
        \tablecolumns{15}
        \tablecaption{Spectral Binary Indices \label{tab:sb}}
        \tablehead{
        \colhead{Name} & \colhead{SpType} & \colhead{SB Class} & \colhead{ref.$^a$} & \colhead{H$_2$O-J} & \colhead{CH$_4$-J} & \colhead{H$_2$O-H} & \colhead{CH$_4$-H} & \colhead{H$_2$O-K} & \colhead{CH$_4$-K} & \colhead{J/K} & \colhead{H-Dip} & \colhead{K-Slope} & \colhead{J-Curve} & \colhead{H-Bump}
        }
        \startdata
CWISE 0205+5944	&	M8	&	---	&	B14	&	1.02	$\pm$	0.00	&	0.77	$\pm$	0.00	&	0.91	$\pm$	0.00	&	1.05	$\pm$	0.00	&	0.48	$\pm$	0.00	&	1.05	$\pm$	0.01	&	0.34	$\pm$	0.00	&	0.99	$\pm$	0.00	&	0.96	$\pm$	0.01	&	0.95	$\pm$	0.00	&	0.94	$\pm$	0.00	\\
CWISE 0224+1526	&	L7	&	---	&	B14	&	0.70	$\pm$	0.02	&	0.73	$\pm$	0.02	&	0.70	$\pm$	0.02	&	1.10	$\pm$	0.02	&	0.38	$\pm$	0.01	&	1.03	$\pm$	0.02	&	0.70	$\pm$	0.01	&	1.01	$\pm$	0.02	&	0.95	$\pm$	0.02	&	1.08	$\pm$	0.03	&	0.83	$\pm$	0.02	\\
CWISE 0227+0830	&	L5 (blue)	&	WEAK	&	B14	&	0.75	$\pm$	0.01	&	0.69	$\pm$	0.01	&	0.70	$\pm$	0.01	&	1.01	$\pm$	0.01	&	0.35	$\pm$	0.00	&	0.92	$\pm$	0.00	&	0.34	$\pm$	0.00	&	0.98	$\pm$	0.00	&	0.93	$\pm$	0.00	&	1.10	$\pm$	0.02	&	0.96	$\pm$	0.01	\\
CWISE 0555+5105	&	M7	&	---	&	B14	&	1.00	$\pm$	0.00	&	0.72	$\pm$	0.00	&	0.90	$\pm$	0.00	&	1.06	$\pm$	0.00	&	0.45	$\pm$	0.00	&	1.09	$\pm$	0.00	&	0.28	$\pm$	0.00	&	0.99	$\pm$	0.00	&	0.93	$\pm$	0.00	&	1.03	$\pm$	0.00	&	0.91	$\pm$	0.00	\\
CWISE 0620+3446	&	L4	&	STRONG	&	B14	&	0.68	$\pm$	0.01	&	0.62	$\pm$	0.01	&	0.73	$\pm$	0.01	&	0.99	$\pm$	0.01	&	0.39	$\pm$	0.01	&	1.00	$\pm$	0.01	&	0.48	$\pm$	0.01	&	0.94	$\pm$	0.01	&	0.93	$\pm$	0.01	&	1.18	$\pm$	0.98	&	0.88	$\pm$	0.01	\\
CWISE 0626+5941	&	L3	&	---	&	B14	&	0.83	$\pm$	0.01	&	0.73	$\pm$	0.00	&	0.74	$\pm$	0.01	&	1.03	$\pm$	0.00	&	0.39	$\pm$	0.00	&	1.01	$\pm$	0.00	&	0.40	$\pm$	0.00	&	0.98	$\pm$	0.00	&	0.91	$\pm$	0.00	&	0.97	$\pm$	0.01	&	0.94	$\pm$	0.00	\\
CWISE 0627-0028	&	T0 (blue)	&	WEAK	&	B10	&	0.66	$\pm$	0.01	&	0.63	$\pm$	0.01	&	0.68	$\pm$	0.01	&	0.87	$\pm$	0.01	&	0.71	$\pm$	0.02	&	0.85	$\pm$	0.02	&	0.37	$\pm$	0.01	&	0.96	$\pm$	0.01	&		---		&		---		&		---		\\
CWISE 0657+1633	&	L6	&	---	&	B14	&	0.73	$\pm$	0.02	&	0.69	$\pm$	0.01	&	0.72	$\pm$	0.01	&	1.10	$\pm$	0.01	&	0.37	$\pm$	0.00	&	0.92	$\pm$	0.00	&	0.64	$\pm$	0.01	&	1.04	$\pm$	0.01	&	0.94	$\pm$	0.00	&	1.14	$\pm$	0.04	&	0.86	$\pm$	0.01	\\
\enddata
        \tablecomments{$^a$ Method used to calculate SB indices: B10 - \cite{2010ApJ...710.1142B}; B14 - \cite{2014ApJ...794..143B}  }
    \end{deluxetable*}
\end{longrotatetable}

\begin{deluxetable*}{cc}
\tabletypesize{\scriptsize}
\tablewidth{0pt}
\tablenum{7}
\tablecolumns{2}
\tablecaption{TESS Rotation Periods \label{table: 5}}
\tablehead{
\colhead{Name} &
\colhead{Period (days)}}
\startdata
TYC 8025-428-1	&	19.5	\\
CWISE J025638.42-335454.9	&	0.8	\\
HD 40781	&	4.6	\\
HD 51400	&	9.8	\\
TYC 8927-2833-1	&	6.5	\\
CWISE J101533.05-111501.0A	&	0.4	\\
WISEA J143934.77-804838.8	&	0.7	\\
UCAC4 840-013771	&	1.3	\\
HD188769	&	1.1	\\
GJ 900	&	11.9	\\
\enddata
\tablecomments{Rotation periods derived via TESS light curves for the observed host stars.}
\end{deluxetable*}

\begin{deluxetable*}{lcccccc}
\tabletypesize{\scriptsize}
\tablewidth{0pt}
\tablenum{8}
\tablecolumns{8}
\tablecaption{Host System Ages \label{table: 6}}
\tablehead{
\colhead{Name} &
\colhead{SpType} &
\colhead{Age$^{a}_{rot}$ (Gyr)} &
\colhead{P$_{rot} ref.$} &
\colhead{Age$^{b}_{Lit}$ (Gyr)} &
\colhead{Lit ref.} &
\colhead{Age$^{c}_{WD}$ (Gyr)}
}
\startdata
WDJ002027.83-153546.50	&	(WD)	&	---	&	---	&	---	&	---	&	6.2$^{+3.9}_{-1.3}$	\\
UCAC3 113-933	&	(M0)	&	---	&	---	&	0.1-1	&	1	&	---	\\
CD-26 134	&	(K3)	&	---	&	---	&	$>$1	&	1	&	---	\\
CD-24 407	&	K0	&	---	&	---	&	$>$1	&	1	&	---	\\
WDJ011001.84-592642.22	&	(WD)	&	---	&	---	&	---	&	---	&	$>$5.9$^{+4.8}_{-3.0}$	\\
WDJ060159.98-462534.40	&	(WD)	&	---	&	---	&	---	&	---	&	10.9$^{+2.6}_{-2.0}$ 	\\
LP 369-22	&	DZ	&	---	&	---	&	---	&	---	&	$>$3.2$^{+0.4}_{-0.3}$	\\
UCAC4 328-061594	&	DC	&	---	&	---	&	---	&	---	&	0.4$\pm$0.1	\\
\enddata
\tablecomments{a- Age range estimated from position on the color-period diagram in comparison to benchmark populations. These ages estimates are not derived from gyrochrones, and should therefore be treated with caution. \\
b- Published age ranges found in the literature. \\
c- Age range calculated using the wddate code. Ages following a ``$>$'' are the WD cooling ages only, not the total age of the system. \\
\textbf{References:} 1-\cite{2017ApJ...835...61Z}; 2-\cite{2019AJ....158...13N}; 3-\cite{2011AA...530A.138C}; 4-\cite{2018ApJ...856...23G}; 5-This paper; 6-\cite{2016ApJ...821...93N}; 7-\cite{2020AA...635A..43R}}
\end{deluxetable*}

\begin{longrotatetable}
\begin{deluxetable*}{lcccccccccccc}
\tabletypesize{\tiny}
\tablewidth{0pt}
\tablenum{9}
\tablecolumns{3}
\tablecaption{Host Star Metallicity. \label{table: 7}}
\tablehead{
\colhead{Name} &
\colhead{[Fe/H]$\_$phot$^a$} &
\colhead{[M/H]$\_$spec$^b$} &
\colhead{[$\alpha$/Fe]$^c$} &
\colhead{[Fe/H]\_{lit}} &
\colhead{Ref$\_$[Fe/H]} &
\colhead{[C/H]} &
\colhead{[O/H]} &
\colhead{[Mg/H]} &
\colhead{[Si/H]} &
\colhead{C/O} &
\colhead{Mg/Si} &
\colhead{Ref.} 
}
\startdata
TYC 8025-428-1	&	0.06	$^{+	0.03	}_{-	0.01	}$	&	-0.07	$^{+	0.03	}_{-	0.04	}$	&	0.10	$^{+	0.02	}_{-	0.02	}$	&		---		&	---	&		---		&		---		&		---		&		---		&	---	&	---	&	---	\\
UCAC3 113-933	&	-0.95	$^{+	0.02	}_{-	0.01	}$	&	-0.22	$^{+	0.10	}_{-	0.32	}$	&	0.06	$^{+	0.05	}_{-	0.02	}$	&	-0.29	$\pm$	0.07	&	1	&		---		&		---		&		---		&		---		&	---	&	---	&	---	\\
WT 5A	&	-0.096	$^{+	0.004	}_{-	0.003	}$	&			---				&			---				&		---		&	---	&		---		&		---		&		---		&		---		&	---	&	---	&	---	\\
CD-26 134	&	0.04	$^{+	0.01	}_{-	0.01	}$	&	0.06	$^{+	0.01	}_{-	0.01	}$	&	0.06	$^{+	0.02	}_{-	0.02	}$	&	0.07	$\pm$	0.07	&	1	&		---		&		---		&		---		&		---		&	---	&	---	&	---	\\
TYC 1744-123-1	&			---				&	-0.03	$^{+	0.09	}_{-	0.11	}$	&	0.13	$^{+	0.02	}_{-	0.02	}$	&		---		&	---	&		---		&		---		&		---		&		---		&	---	&	---	&	---	\\
LAMOST J004950.95+423653.6	&			---				&	-0.43	$^{+	0.25	}_{-	0.27	}$	&	0.14	$^{+	0.07	}_{-	0.09	}$	&	-1.39	$\pm$	0.06	&	2	&		---		&		---		&		---		&		---		&	---	&	---	&	---	\\
CD-24 407	&	-0.12	$^{+	0.01	}_{-	0.01	}$	&			---				&			---				&		---		&	---	&		---		&		---		&		---		&		---		&	---	&	---	&	---	\\
LP 883-372	&	-1.34	$^{+	0.00	}_{-	0.00	}$	&	-0.14	$^{+	0.3	}_{-	0.05	}$	&	-0.02	$^{+	0.02	}_{-	0.03	}$	&		---		&	---	&		---		&		---		&		---		&		---		&	---	&	---	&	---	\\
\enddata
\tablecomments{a- [Fe/H] from the Gaia DR3 GSP-Phot Aeneas best library. \\ b- [M/H] from the Gaia DR3 GSP-Spec MatisseGauguin. \\ c- Abundance of $\alpha$ elements from the Gaia DR3 GSP-Spec MatisseGauguin. \\
\textbf{References}: DR3- Gaia DR3; 1-\cite{2017AJ....153...75K}; 2-\cite{2019ApJS..245...34X}; 3-\cite{2019yCat.5164....0L}; 4-\cite{2022ApJS..260...45D}; 5-\cite{2016ApJ...826..171G}; 6-\cite{2022AA...659A..95T}; 7-\cite{2022AJ....163..152S}; 8-\cite{2014MNRAS.443.2561G}; 9-\cite{2020ApJS..251...15Z}; 10-\cite{2022AA...660A..38W}; 11-\cite{2022ApJS..261...36X}; 12-\cite{2014AJ....148...54H}; 13-\cite{2011AJ....142..193B}; 14-\cite{2020ApJ...891...23W} }
\end{deluxetable*} 
\end{longrotatetable}

\startlongtable
\begin{deluxetable*}{llcccccc}
\tabletypesize{\tiny}
\tablewidth{0pt}
\tablenum{10}
\tablecolumns{8}
\tablecaption{Binding Energies \label{table: 9}}
\tablehead{
\colhead{Host} &
\colhead{Companion} &
\colhead{M$_{Host}$\tablenotemark{a}} &
\colhead{M$_{Comp}$ } &
\colhead{Total Mass } &
\colhead{Mass Ratio} &
\colhead{Binding Energy} &
\colhead{$\tau$\tablenotemark{b}} \\
\colhead{} &
\colhead{} &
\colhead{($M_{\sun}$)} &
\colhead{($M_{\sun}$)} &
\colhead{($M_{\sun}$)} &
\colhead{} &
\colhead{(10$^{41}$ erg)} &
\colhead{(Gyrs)}
}
\startdata
TYC 8025-428-1	&	CWISE J000021.45-481314.9	&	0.89	&	0.06	&	0.95	&	0.06	&	1.0	&	$>$14	\\
WDJ002027.83-153546.50	&	CWISE J002029.72-153527.5	&	0.64	&	0.08	&	0.71	&	0.12	&	2.0	&	$>$14	\\
UCAC3 113-933	&	CWISE J002101.45-334631.5	&	0.61	&	0.02	&	0.63	&	0.03	&	0.30	&	$>$14	\\
2MASS J00215427-4514590	&	CWISE J002159.02-451434.4	&	0.15	&	0.03	&	0.18	&	0.23	&	0.30	&	13	\\
WT 5AB	&	CWISE J002414.30-403053.7	&	0.37	&	0.06	&	0.42	&	0.15	&	0.70	&	$>$14	\\
CD-26 134	&	CWISE J002658.73-260859.6	&	0.80	&	0.06	&	0.86	&	0.07	&	2.0	&	$>$14	\\
TYC 1744-123-1	&	CWISE J004218.67+291730.6	&	0.63	&	0.06	&	0.69	&	0.09	&	0.70	&	$>$14	\\
LAMOST J004950.95+423653.6	&	CWISE J004945.38+423619.3	&	0.37	&	0.06	&	0.43	&	0.15	&	0.70	&	$>$14	\\
\enddata
\tablecomments{Summary of the masses, mass ratios (M$_{Companion}$/M$_{Host}$), total Masses and binding energies for the sample.}
\tablenotetext{a}{Host mass includes the masses of all known host system components, i.e. both components A and B if the host were binary.}
\tablenotetext{b}{Estimated dissipation lifetime of the system, calculated as described in Section \ref{sec:be}.}
\end{deluxetable*}

\begin{deluxetable*}{cccccccc}
\tabletypesize{\tiny}
\tablewidth{0pt}
\tablenum{11}
\tablecolumns{8}
\tablecaption{Host System Multiplicity \label{table: 8}}
\tablehead{
 \colhead{\textbf{Confirmed}} & & & & & & & \\
\colhead{Name 1} &
\colhead{Name 2} &
\colhead{SpType 1} &
\colhead{Sptype 2} &
\colhead{Plx 1 (mas)} &
\colhead{Plx 2 (mas)} &
\colhead{Sep. ($\arcsec$)} &
\colhead{Sep. (AU)}
}
\startdata
WT 5A	&	WT 5B	&	(M4.5)	&	(M4.5)	&	18.66$\pm$0.07	&	18.56$\pm$0.03	&	1.8	&	10	\\
LSPM J0407+1911A	&	LSPM J0407+1911B	&	(M4)	&	(M4)	&	18.77$\pm$0.06	&	18.90$\pm$0.03	&	2	&	110	\\
G 194-47A	&	G 194-47B	&	M3.5	&	(M3.5)	&	55.73$\pm$0.42	&	---	&	0.3	&	5	\\
StKM 1-760	&	LP 369-22	&	K6	&	DZ	&	20.61$\pm$0.80	&	21.44$\pm$0.13	&	44	&	2,100	\\
CWISE J101533.05-111501.0A	&	CWISE J101533.05-111501.0B	&	M5	&	(M5)	&	23.95$\pm$0.03	&	24.14$\pm$0.03	&	2.6	&	110	\\
BD+49 2561A & BD+49 2561B & K3 & M2 & 17.53$\pm$0.01 & 17.06$\pm$0.16 & 15 & 860 \\
HD188769	&	DENIS J200057.2-644827	&	F3IV	&	M5	&	20.40$\pm$0.51	&	24.23$\pm$0.06	&	22.6	&	930	\\
LSPM J2214+2534A	&	LSPM J2214+2534B	&	M4.3	&	M4.3	&	24.25$\pm$0.49	&	---	&	0.1	&	4	\\
\enddata
\tablecomments{Top: List of host stars in the sample which are confirmed binary or multiple systems. Bottom: List of host stars in the sample which are suspected binaries. \\
$*$ - Separations between the A-BC and B-C components respectively. \\
a: RUWE value from Gaia DR2. \\
b: RUWE Value from Gaia DR3. \\
c: Denotes whether or not the source appears in a catalog suspecting binarity. 1 - Gaia DR3 nss$\_$two$\_$body$\_$orbit; 2- Gaia DR3 Performance Verification: binary$\_$masses; 3- Gaia DR3 nss$\_$acceleration$\_$astro; 4- \cite{2014AJ....147...86T}. \\
d: From \cite{2022AA...664A.105F} \\
e: From \cite{2017ApJS..230...24B} \\
$**$- Data taken from \cite{2018AA...614A.125F}
}
\end{deluxetable*}

\clearpage

\bibliography{byw}
\bibliographystyle{aasjournal}
\end{document}